\documentclass[12pt]{article}
\usepackage{epsfig}

\usepackage{mathrsfs}
\usepackage[T1]{fontenc}
\usepackage{mathpazo}
\usepackage{setspace}
\usepackage{amsfonts}
\usepackage{amssymb}
\usepackage{amsmath}
\usepackage{epsfig}
\usepackage{latexsym}
\usepackage{color}
\usepackage{graphicx}
\usepackage{nicefrac}
\usepackage[latin1]{inputenc}
\usepackage{pstricks}
\usepackage{slashed}

\def\hybrid{\topmargin -20pt    \oddsidemargin 0pt
        \headheight 0pt \headsep 0pt
        \textwidth 6.25in       
        \textheight 9.5in       
        \marginparwidth .875in
        \parskip 5pt plus 1pt   \jot = 1.5ex}

\hybrid

\catcode`\@=11

\def\marginnote#1{}
\newcount\hour
\newcount\minute
\newtoks\amorpm
\hour=\time\divide\hour by60 \minute=\time{\multiply\hour by60
\global\advance\minute by-\hour}
\edef\standardtime{{\ifnum\hour<12 \global\amorpm={am}%
        \else\global\amorpm={pm}\advance\hour by-12 \fi
        \ifnum\hour=0 \hour=12 \fi
        \number\hour:\ifnum\minute<10 0\fi\number\minute\the\amorpm}}
\edef\militarytime{\number\hour:\ifnum\minute<10
0\fi\number\minute}

\def\draftlabel#1{{\@bsphack\if@filesw {\let\thepage\relax
   \xdef\@gtempa{\write\@auxout{\string
      \newlabel{#1}{{\@currentlabel}{\thepage}}}}}\@gtempa
   \if@nobreak \ifvmode\nobreak\fi\fi\fi\@esphack}
        \gdef\@eqnlabel{#1}}
\def\@eqnlabel{}
\def\@vacuum{}
\def\draftmarginnote#1{\marginpar{\raggedright\scriptsize\tt#1}}

\def\draft{\oddsidemargin -.5truein
        \def\@oddfoot{\sl preliminary draft \hfil
        \rm\thepage\hfil\sl\today\quad\militarytime}
        \let\@evenfoot\@oddfoot \overfullrule 3pt
        \let\label=\draftlabel
        \let\marginnote=\draftmarginnote
   \def\@eqnnum{(\theequation)\rlap{\kern\marginparsep\tt\@eqnlabel}%
\global\let\@eqnlabel\@vacuum}  }

\def\preprint{\twocolumn\sloppy\flushbottom\parindent 2em
        \leftmargini 2em\leftmarginv .5em\leftmarginvi .5em
        \oddsidemargin -.5in    \evensidemargin -.5in
        \columnsep .4in \footheight 0pt
        \textwidth 10.in        \topmargin  -.4in
        \headheight 12pt \topskip .4in
        \textheight 6.9in \footskip 0pt
        \def\@oddhead{\thepage\hfil\addtocounter{page}{1}\thepage}
        \let\@evenhead\@oddhead \def\@oddfoot{} \def\@evenfoot{} }

\def\numberbysection{\@addtoreset{equation}{section}
        \def\theequation{\thesection.\arabic{equation}}}

\def\underline#1{\relax\ifmmode\@@underline#1\else
        $\@@underline{\hbox{#1}}$\relax\fi}

\def\titlepage{\@restonecolfalse\if@twocolumn\@restonecoltrue\onecolumn
     \else \newpage \fi \thispagestyle{empty}\c@page\z@
        \def\thefootnote{\fnsymbol{footnote}} }

\def\endtitlepage{\if@restonecol\twocolumn \else \newpage \fi
        \def\thefootnote{\arabic{footnote}}
        \setcounter{footnote}{0}}  

\catcode`@=12 \relax

\def\figcap{\section*{Figure Captions\markboth
        {FIGURECAPTIONS}{FIGURECAPTIONS}}\list
        {Figure \arabic{enumi}:\hfill}{\settowidth\labelwidth{Figure
999:}
        \leftmargin\labelwidth
        \advance\leftmargin\labelsep\usecounter{enumi}}}
 \relax
\def\tablecap{\section*{Table Captions\markboth
        {TABLECAPTIONS}{TABLECAPTIONS}}\list
        {Table \arabic{enumi}:\hfill}{\settowidth\labelwidth{Table
999:}
        \leftmargin\labelwidth
        \advance\leftmargin\labelsep\usecounter{enumi}}}
 \relax
\def\reflist{\section*{References\markboth
        {REFLIST}{REFLIST}}\list
        {[\arabic{enumi}]\hfill}{\settowidth\labelwidth{[999]}
        \leftmargin\labelwidth
        \advance\leftmargin\labelsep\usecounter{enumi}}}
 \relax
\makeatletter
\newcounter{pubctr}
\def\publist{\@ifnextchar[{\@publist}{\@@publist}}
\def\@publist[#1]{\list
        {[\arabic{pubctr}]\hfill}{\settowidth\labelwidth{[999]}
        \leftmargin\labelwidth
        \advance\leftmargin\labelsep
        \@nmbrlisttrue\def\@listctr{pubctr}
        \setcounter{pubctr}{#1}\addtocounter{pubctr}{-1}}}
\def\@@publist{\list
        {[\arabic{pubctr}]\hfill}{\settowidth\labelwidth{[999]}
        \leftmargin\labelwidth
        \advance\leftmargin\labelsep
        \@nmbrlisttrue\def\@listctr{pubctr}}}
 \relax
\makeatother
\newskip\humongous \humongous=0pt plus 1000pt minus 1000pt

\newif\ifdtup

\relax

\def\be{\begin{equation}}
\def\ee{\end{equation}}
\def\ba{\begin{eqnarray}}
\def\ea{\end{eqnarray}}

\renewcommand{\theequation}{\thesection.\arabic{equation}}

\newcommand{\eqn}[1]{(\ref{#1})}

\author{
  \begin{minipage}{.97\linewidth}
    \vspace{1cm}
    \begin{center}
      \begin{small}
        \textbf{Ioannis Bakas}\footnote{bakas@mail.ntua.gr} ${\ }^1$ and
               \textbf{Dieter L\"ust}\footnote{dieter.luest@lmu.de} ${\ }^{2,3,4}$
        \end{small}
    \end{center}
    \vspace{0.5cm}
    \hspace{1.5cm}\begin{minipage}{.8\linewidth}
     {\it \begin{footnotesize}
    \begin{itemize}
        \item[${}^1$] Department of Physics, School of Applied Mathematics and Physical Sciences \\
        National Technical University, 15780 Athens, Greece
      \item[${}^2$] Max-Planck-Institut f\"ur Physik\\
       F\"ohringer Ring 6, 80805 M\"unchen, Germany
                    \item[${}^3$] Arnold-Sommerfeld-Center f\"ur Theoretische Physik\\
        Department f\"ur Physik, Ludwig-Maximilians-Universit\"at M\"unchen\\
        Theresienstra\ss e 37, 80333 M\"unchen, Germany
        \item[${}^4$] Theory Group, Department of Physics, CERN\\
        CH-1211 Geneva 23, Switzerland
       \end{itemize}
     \end{footnotesize}}
    \end{minipage}
    \vspace{0.5cm}
  \end{minipage}
}

\title{\vspace{1.6cm}
 \boldmath \begin{LARGE}
    \textbf{T-duality, Quotients and Currents for \\
    Non-Geometric Closed Strings}
  \end{LARGE} \unboldmath
}

\begin{document}

\renewcommand{\thepage}{\arabic{page}}
\setcounter{page}{1}


\begin{titlepage}
  \maketitle
  \thispagestyle{empty}

  \vspace{-16.9cm}
  \begin{flushright}
    LMU-ASC 27/15\\
    MPP-2015-96\\
    CERN-PH-TH-2015-103
  \end{flushright}

  \vspace{14cm}

  \begin{center}
    \textsc{Abstract}\\
  \end{center}
 We use the canonical description of T-duality as well as the formulation of T-duality in terms of chiral currents
 to investigate the geometric and non-geometric faces of closed
 string backgrounds originating from principal torus bundles with constant H-flux.  Employing   conformal field theory
 techniques, the  non-commutative and non-associative structures among generalized coordinates in the so called Q-flux and R-flux
 backgrounds emerge by gauging the Abelian symmetries  of an enlarged Ro\v{c}ek-Verlinde sigma-model and projecting the
 associated chiral currents of the enlarged theory to the
 T-dual coset models carrying non-geometric fluxes.
\end{titlepage}

\tableofcontents


\section{Introduction}
\setcounter{equation}{0}

Flux vacua of closed string theory have been investigated in great detail in recent years (for 
topical reviews, see, for instance,  \cite{Grana:2005jc,Blumenhagen:2006ci}). Various superstring backgrounds
were brought to light in that context, including examples of geometric as well as non-geometric spaces.
Non-geometry is a  relatively new concept in string theory whose necessity and importance were 
unveiled in a series of papers by employing T-duality and its coordinate
dependent generalizations
\cite{Hellerman:2002ax}-\cite{Andriot:2013xca}, but it should also be noted that non-geometric string constructions 
appeared long time ago in conformal field theory using free fermions,
free bosons and asymmetric orbifolds \cite{Kawai:1986ah}-\cite{Narain:1986qm}.
It was found that closed strings on non-geometric flux backgrounds have surprising features in that they exhibit
non-commutativity and non-associativity among the closed string coordinates and their momenta
\cite{Blumenhagen:2010hj,Lust:2010iy}. These new algebraic structures were investigated further
in a subsequent series of papers \cite{Blumenhagen:2011ph}-\cite{Aschieri:2015roa},
focusing on the derivation of the commutation relations for different string models and their mathematical 
interpretation. Independent studies by other authors, following a more mathematical line of thought,
have also come to the conclusion that non-commutative and
non-associative tori arise from geometric flux vacua by a sequence of T-dualities \cite{mathai, bouwknegt4}.

Non-geometric backgrounds capture properties of string theory that cannot be formulated in terms of
conventional compactifications of supergravity. The prime example is provided by the T-dual faces of
toroidal flux vacua, known as $Q$- and $R$-flux models, but there are also examples of non-geometric spaces that do
not have geometric duals  \cite{Condeescu:2013yma,Hassler:2014sba}.
However, it remains a challenge to this day how to define and explore the properties of non-geometric closed string vacua
in all generality and find a unifying principle to describe geometry and non-geometry on equal footing. Double geometry and
double field theory (DFT) seem to be the most prominent candidates for encompassing string models and their T-dual
faces in a democratic way  \cite{Tseytlin:1990nb}-\cite{Blumenhagen:2015zma}.

There are two different classes of non-geometric spaces. The first class are $Q$-spaces, which are locally geometric in
that they resemble differentiable manifolds locally, but they fail to be globally geometric because their
gluing conditions involve T-duality transformations and not just diffeomorphisms. In that case,
departure from ordinary geometry is mild and it manifests as non-commutativity of the closed string coordinates.
The target space still makes good sense as a fibration though the monodromy of the fibres, as one goes around the
base space, is non-trivial. The second class are $R$-spaces, which fail to be differentiable manifolds globally as
well as locally. Then, the concept of target space geometry breaks down completely and the string
coordinates exhibit commutation relations of a non-commutative and non-associative algebra.

It is the aim of the present work to provide an alternative conformal field theory (CFT) derivation of the commutation relations
among the closed string coordinates of non-geometric flux vacua, focusing attention to the well known class of three-dimensional 
models with parabolic monodromies, which carry $H$-, $f$-, $Q$- and $R$-flux and they are interrelated by T-dualities.
The commutation relations of the closed string coordinates of these models have already been derived using the  operator approach  \cite{Andriot:2012vb} (but see also the computation based on Dirac brackets that was reported in ref. \cite{Blair:2014kla}). 
The approach we take here is entirely based on CFT world-sheet methods and it complements
nicely the previous work on the subject that brought to light the non-commutative and non-associative aspects of
those flux string models.  The present paper also extends and refines the CFT computations of ref. \cite{Blumenhagen:2011ph}, where 
the non-associativity was derived from the structure of CFT scattering amplitudes among tachyon vertex operators in the $R$-flux 
background. The results of ref. \cite{Blumenhagen:2011ph} led to the introduction of a non-associative tri-product among functions,
which was further investigated in \cite{Mylonas:2012pg,Bakas:2013jwa}.
However, as noted in \cite{Blumenhagen:2011ph}, the
non-associative structure of CFT scattering amplitudes in the $R$-flux background is only visible when some of the tachyon momenta 
are taken off-shell, whereas on-shell the
physical scattering amplitudes are associative, as expected in any conformal field theory. This result was analyzed further in 
the context of double field theory \cite{Blumenhagen:2013zpa}, showing, in particular, that the non-associative tri-product acts 
non-trivially on functions of doubled space only when the strong DFT constraint is violated.

In this paper, we compute directly the underlying commutation relations
of some suitably defined {\sl generalized} string coordinates of the $H$-, $f$-, $Q$- and $R$-flux backgrounds, 
without making reference to off-shell nor 
on-shell amplitudes. For this purpose, we start from the ungauged non-linear sigma-model of Ro\v{c}ek and Verlinde
\cite{Rocek:1991ps}, which provides the "parent" theory in our framework. It contains chirally conserved $U(1)$ world-sheet currents
${\cal J}^I$ and $\bar{\cal  J}^I$, which transform in a simple linear way under T-duality transformations.
The associated T-dual "child" theories are obtained by gauging the associated axial/vector like $U(1)$ symmetries.  
The $U(1)$ currents ${\cal J}^I$ and $\bar{\cal  J}^I$ are not anymore chirally conserved when they are expressed in 
terms of the target space coordinates $X^I$ of the associated "child" theories, but they still
transform linearly under T-duality. Actually, as will be seen later in detail, the non-commutative and/or non-associative  
structures that arise in the presence of non-geometric fluxes refer to generalized string coordinates ${\cal X}^I$, 
which are defined by the world-sheet integrals of the Ro\v{c}ek-Verlinde currents ${\cal J}^I$ and $\bar{\cal  J}^I$ as 
${\cal X}^I=\int {\cal J}^Idz$ and $\bar {\cal X}^I=\int \bar {\cal J}^Id\bar z$ with prescribed integration paths.
The generalized coordinates also transform nicely under T-duality, mapping the commutative geometry of geometric flux backgrounds 
to the non-commutative geometry of the non-geometric flux backgrounds via T-duality, as it has already been advocated in 
ref. \cite{Lust:2010iy}. 

It should be noted at this point that the currents ${\cal J}^I$ and $\bar{\cal  J}^I$ used in this 
paper are refined compared to the choice of currents made in ref. \cite{Blumenhagen:2011ph}.
It should also be emphasized that the non-commutative and/or non-associative algebras that we are discussing here do not 
arise among the sigma-model coordinates $X^I$ of the child theories with non-geometric fluxes, but they rather emerge
by projecting the chiral currents of the parent theory to the child gauged sigma-models and expressing the 
generalized coordinates ${\cal X}^I$ as world-sheet integrals of those non-chiral currents. We further note that the 
algebra of currents in the parent Ro\v{c}ek-Verlinde theory is completely commutative and associative, unlike the non-trivial 
commutation relations that can emerge in the quotient. All these put on firm basis the existing results for toroidal flux models
and they also provide the means to explore generalizations to other non-geometric backgrounds of current interest, including  
the $Q$- and $R$-frames of Wess-Zumino-Witten models on group manifolds (see,
for instance, \cite{schulz,Blumenhagen:2014gva,Blumenhagen:2015zma}).

The material of this paper is organized as follows. In section 2, we recall the description of T-duality as canonical
transformation \cite{Buscher:1987sk}-\cite{Giveon:1994}
on the coordinates $X^I$ and momenta $P_I$ of the world-sheet sigma-model, hereby introducing the notion of dual
coordinates $\tilde X^I$ and dual momenta $\tilde P^I$, which, in general, are expressed non-locally in terms of the original phase space
variables. We also consider the chiral currents of the parent Ro\v{c}ek-Verlinde model
and combine them with the canonical approach to T-duality, hereby expressing those currents in terms of the (dual) coordinates 
of the gauged child theories. The action of T-duality as automorphism of the currents in the underlying
conformal field theory turns out to be useful for string backgrounds without isometries, giving rise to non-geometries.
In section 3, we focus on toroidal flux models originating from $T^3$ with constant $H$-flux and explain how 
T-duality is used at work to obtain the geometric and non-geometric faces of closed string spaces.
We also define the generalized coordinates ${\cal X}^I$ of the parabolic flux backgrounds via the projected 
currents of the parent Ro\v{c}ek-Verlinde theory. In section 4, we confirm the emergence
of non-commutative and non-associative structures in the commutation relations of the generalized closed strings coordinates
and momenta of the $Q$-flux and $R$-flux models, respectively, by conformal field theory methods.
Finally, in section 5, we present our conclusions and discuss some open problems for future work. There are also three 
appendices summarizing the basic definitions and properties of the monodromies arising in $T^2$ fibrations over $S^1$ (Appendix A),
the salient features of the parabolic $H$- $f$- $Q$- and $R$-flux models, which are detached from the main text (Appendix B), 
as well as some dilogarithmic integrals, based on the definitions and properties of the dilogarithm function found in
ref. \cite{zagier}, which are repeatedly used in the text for the computation of the commutation relations (Appendix C).

Certain aspects of the canonical approach to string non-commutativity were also discussed recently in ref. \cite{Nikolic},
following a complementary route. There, the monodromies of toroidal fibrations are not built into the closed string
boundary conditions a priori, but they rather come out a posteriori. In this sense, there is partial overlap of their results
with our section 4, but our presentation is more natural for the purposes of string theory. Our work also emphasizes for the first 
time the use of generalized string coordinates.

\section{Canonical T-duality and redefined world-sheet currents}
\setcounter{equation}{0}

Our starting point is the world-sheet action of a general non-linear sigma-model with metric $G$ and anti-symmetric
tensor field $B$ couplings that provide the background for string propagation,
\begin{equation}\label{wsaction}
S={1 \over 2\pi \alpha^{\prime}} \int  dz d \bar{z} \left( G_{IJ}(X) + B_{IJ}(X) \right) \partial X^I \bar{\partial} X^J \ .
\end{equation}
The inverse string tension is $\alpha^{\prime} \sim l_s^2$ and it can be normalized to $1$.
The target space coordinates $X^I$ are two-dimensional fields labeled by $I = 1, \, 2, \, \cdots , \, d$.
The action \eqn{wsaction} is often used to describe the entire space in which the string propagates, but it can also
be part of it when it is tensored together with other blocks to maintain conformal invariance upon quantization.
There is also a dilaton field $\Phi (X)$ that enters in the renormalization group analysis of the sigma-model,
but this is suppressed here.

In this section, we review the T-duality rules of the sigma-model in the presence of an isometry in target
space \cite{Buscher:1987sk} and
use them to discuss the general features of the commutation relations among the coordinates in the T-dual faces of
string backgrounds. Explicit calculations will be performed in subsequent sections for specific models.
We focus on the interpretation of T-duality as canonical transformation in the phase space of the two-dimensional
sigma-model \eqn{wsaction} \cite{amit, Alvarez:1994wj, Bakas:1995hc}, as well as on the axial-vector quotient description
of T-dual backgrounds that follows by gauging chiral symmetries \cite{Rocek:1991ps}. These two approaches are
complementary to each, leading to the important notion of dual coordinates. The dual coordinates are expressed
non-locally in terms of the original fields, accounting for non-commutativity/non-associativity in the applications that
will be considered later.

Note that the duality rules are applicable to all sigma-models, including, for example, principal chiral models,
irrespective of conformal invariance. However, if the original model is conformal, the
dual model will also be conformal for appropriate choice of the dilaton field. In the latter case, T-duality
acts as a solution generating symmetry at the fixed points of the renormalization group equations and it is
promoted to an exact symmetry of string theory relating two conformal field theories with different target
space geometries.
Further details and applications of target space duality in string theory can be found in the report \cite{Giveon:1994},
and references therein, which also discuss the continuous T-duality rules and their discrete variants for general
backgrounds admitting $n$ commuting isometries in terms of $O(n,n)$ group elements.

\subsection{T-duality as canonical transformation}

Suppose that the background does not depend on the coordinate $X^1$, i.e., $\partial / \partial X^1$ is an isometry of the world-sheet action
in adapted coordinates of target space. The components of the corresponding Killing vector field $\xi = k^I \partial / \partial X^I$
are $k^I = (1, 0, \cdots , 0)$ and $k^2 = G_{IJ} k^I k^J$ is the length-squared of $\xi$.
Setting $z = (\tau + \sigma)/2$ and $\bar{z} = (\tau - \sigma)/2$, so that
$\partial = \partial / \partial \tau + \partial / \partial \sigma$ and $\bar{\partial} = \partial / \partial \tau -
\partial / \partial \sigma$, the Lagrangian density takes the following form
\be
{\cal L} = {V \over 2} \left((\dot{X}^1)^2 - ({X^1}^{\prime})^2 \right) + \dot{X}^1 (J_1 + \bar{J}_1) -
{X^1}^{\prime} (J_1 - \bar{J}_1) + U ~,
\ee
where
\be
V = G_{11} = k^2 ~, ~~~~~~ U = {1 \over 2} \left(G_{ij} + B_{ij}\right) \partial X^i ~ \bar{\partial} X^j
\ee
with $i$ and $j$ taking all other values of $I$ and $J$ but $1$ and
\be
J_1 (z, \bar{z}) = {1 \over 2} \left(G_{1i} - B_{1i}\right) \partial X^i ~ , ~~~~~~
\bar{J}_1 (z, \bar{z}) = {1 \over 2} \left(G_{1i} + B_{1i}\right) \bar{\partial} X^i ~.
\ee

Next, we apply Legendre transformation to the pair of variables $(X^1, \dot{X}^1)$. The momentum conjugate to the
coordinate $X^1$ is
\be
P_1 = V \dot{X}^1 + J_1 + \bar{J}_1
\label{momentara1}
\ee
and, hence, the Hamiltonian density (or better to say the {\em Routhian}, since the transformation is performed only with respect
to the cyclic coordinate $X^1$ that does not appear explicitly in the Lagrangian) is
\be
{\cal H} = {1 \over 2V} (P_1)^2 + {V \over 2}({X^1}^{\prime})^2 + {1 \over 2V} (J_1 + \bar{J}_1)^2 -
{1 \over V} P_1 (J_1 + \bar{J}_1) + {X^1}^{\prime} (J_1 - \bar{J}_1) - U ~.
\ee
Let us now perform a canonical transformation $(X^1 , P_1) \rightarrow (\tilde{X}_1 , \tilde{P}^1)$
generated by the function
\be
{\cal F} = {1 \over 2} \oint_{S^1}
(X^1 \tilde{X}_1^{\prime} - {X^1}^{\prime} \tilde{X}_1) d \sigma
\label{genafunc}
\ee
so that
\be
P_1 = {\delta {\cal F} \over \delta X^1} = \tilde{X}_1^{\prime} ~, ~~~~~~
\tilde{P}^1 = - {\delta {\cal F} \over \delta \tilde{X}_1} = {X^1}^{\prime} ~.
\label{loustros}
\ee
We call $\tilde{X}_1$ and $\tilde{P}^1$ the dual coordinate and momentum to $X^1$ and $P_1$, respectively. Then,
in terms of the dual variables, the Hamiltonian takes the form
\be
\tilde{{\cal H}} = {V \over 2} (\tilde{P}^1)^2 + {1 \over 2V}(\tilde{X}_1^{\prime})^2 + {1 \over 2V} (J_1 + \bar{J}_1)^2
- {1 \over V} \tilde{X}_1^{\prime} (J_1 + \bar{J}_1) + \tilde{P}^1 (J_1 - \bar{J}_1) - U ~.
\ee
Varying $\tilde{{\cal H}}$ with respect to $\tilde{P}^1$, we find that the dual velocity and momentum are related to
each other as follows,
\be
\dot{\tilde{X}}_1 = V \tilde{P}^1 + J_1 - \bar{J}_1 ~.
\label{momentara2}
\ee
Finally, we perform the inverse Legendre transformation with respect to $(\tilde{X}_1 , \tilde{P}^1)$ and arrive
at the dual Lagrangian
\be
\tilde{{\cal L}} = {1 \over 2V} \left((\dot{\tilde{X}}_1)^2 - (\tilde{X}_1^{\prime})^2 \right) - {1 \over V}
\dot{\tilde{X}}_1 (J_1 - \bar{J}_1) + {1 \over V} \tilde{X}_1^{\prime} (J_1 + \bar{J}_1) + \tilde{U} ~,
\ee
where
\be
\tilde{U} = U - {2 \over V} J_1 \bar{J}_1 ~.
\ee

The form of the dual Lagrangian $\tilde{{\cal L}}$ is identical to the original Lagrangian ${\cal L}$ provided that
the background fields of the dual non-linear sigma-model with target space coordinates $(\tilde{X}_1, X^i)$ are taken to be
\ba
& & \tilde{G}_{11} = {1 \over G_{11}} ~, ~~~~ \tilde{G}_{1i} = {B_{1i} \over G_{11}} ~, ~~~~
\tilde{G}_{ij} = G_{ij} - {G_{1i} G_{1j} - B_{1i} B_{1j} \over G_{11}} ~, \nonumber\\
& & \tilde{B}_{1i} = {G_{1i} \over G_{11}} ~, ~~~~
\tilde{B}_{ij} = B_{ij} - {G_{1i} B_{1j} - B_{1i} G_{1j} \over G_{11}} ~.
\label{toutaforma}
\ea
These are precisely Buscher rules for T-duality with respect to a Killing vector field $\xi$ in adapted
coordinates \cite{Buscher:1987sk}, which are formulated as canonical transformation in the phase space of the two-dimensional
non-linear sigma-model \cite{amit, Alvarez:1994wj, Bakas:1995hc}. A more
compact way to describe the T-duality transformation along $X^1$ is provided by combining the metric and anti-symmetric tensor
fields as $E_{IJ}=G_{IJ}+B_{IJ}$ and letting
\begin{eqnarray}\label{tdualgb}
\tilde{E}_{1 1}=\frac{1}{E_{11}}\, , ~~~~
\tilde{E}_{1 i}=\frac{E_{1i}}{E_{11}}\, , ~~~~
\tilde{E}_{i 1}=-\frac{E_{i 1}}{E_{11}}\, , ~~~~
\tilde{E}_{ij}=E_{ij}-\frac{E_{i1 } E_{1 j}}{E_{11}}\, .\label{BuscherE}
\end{eqnarray}
Conformal invariance of the world-sheet sigma-model also requires that the corresponding dilaton field transforms as
$\tilde{\Phi} = \Phi -\log V$.

The effect of T-duality can be neatly described by a non-local redefinition of the target space coordinate associated to the
Killing isometry. For this purpose, we write down the transformation of the derivatives of the Killing coordinate
\begin{eqnarray}\label{wstdual}
 \partial X^1 ={1\over E_{11}}(\partial \tilde{X}_1-E_{i1}\partial X^i) \, , ~~~~~
  \bar\partial X^1 =-{1\over E_{11}}(\bar\partial \tilde{X}_1+E_{1i}\bar\partial X^i)
\end{eqnarray}
by combining the canonical transformation \eqn{loustros} with the expressions for the momenta \eqn{momentara1} and \eqn{momentara2}.
Substituting the expressions \eqn{BuscherE}, we arrive at the integral formula
\begin{equation}\label{dualcoordinates}
X^1(z,\bar z)=\int ^z (\tilde{E}_{11} \partial \tilde{X}_1 + \tilde{E}_{i1}\partial X^i)  dz' -\int ^{\bar z}
(\tilde{E}_{11} \bar\partial \tilde{X}_1 + \tilde{E}_{1i}\bar\partial X^i) d\bar z'
\end{equation}
expressing $X^1$ in terms of the dual background.
There is an analogous formula expressing $\tilde{X}_1$ in terms of the original background, as
\be
\partial \tilde{X}_1 = E_{11} \partial X^1 + E_{i1}\partial X^i \, , ~~~~~
\bar\partial \tilde{X}_1 = - (E_{11} \bar\partial X^1+E_{1i}\bar\partial X^i) ~,
\label{ninlaude}
\ee
which integrates to
\begin{equation}
\tilde{X}_1(z,\bar z)=\int ^z (E_{11} \partial X^1 + E_{i1}\partial X^i)  dz' -\int ^{\bar z} (E_{11}
\bar\partial X^1+E_{1i}\bar\partial X^i) d\bar z'\,.
\label{tsitsiol}
\end{equation}
These expressions show that T-duality acts, in general, in a rather complicated {\em non-local} way on the target space
coordinates.

For constant backgrounds, i.e., for a free two-dimensional conformal field theory with $E_{IJ}=\delta_{IJ}$, the action of
T-duality looks very simple. For compactifications on a circle, the left- and right-moving coordinates are chiral,
\be
X_L^1 (z) = \int^z   \partial X^1 (z^{\prime}, \bar{z}^{\prime}) dz^{\prime} ~, ~~~~~~ X_R^1 (\bar{z}) = \int^{\bar{z}}
\bar\partial X^1 (z^{\prime}, \bar{z}^{\prime}) d\bar{z}^{\prime} ~,
\ee
and they transform as follows, by specializing equation \eqn{tsitsiol},
\begin{equation}\label{freetrans}
  \tilde{{X}}_{1L} = X^1_L\, ,\quad \tilde{{X}}_{1R} =- X^1_R\, .
  \end{equation}
Hence, one immediately obtains the well-known T-duality relation among the coordinates $X^1$ and $\tilde X_1$
of the two dual models,
\begin{equation}
X^1=X^1_L+X^1_R\,\, \longrightarrow\,\,  \tilde X_1=X_L^1-X_R^1\, .
\label{netaeffa}
\end{equation}

In all other cases, one can still introduce variables $X^1_L$ and $X^1_R$ by adding and subtracting the integral
expressions \eqn{dualcoordinates} and \eqn{tsitsiol}, thus decomposing the coordinates $X^1$ and $\tilde{X}_1$ as
\be
X^1=X^1_L+X^1_R \, , ~~~~~~ \tilde X_1=X_L^1-X_R^1 ~.
\ee
The effect of T-duality is still summarized by equation \eqn{netaeffa}, as for the case of free fields. Note, however,
that $X_L^1$ and $X_R^1$ are not chiral and anti-chiral world-sheet functions, in general, because
$\partial X^1 (z,\bar z)$ and $\bar\partial X^1 (z,\bar z)$ are not separately conserved on general backgrounds.
Although the general case of non-constant background fields $E_{IJ}(X^i)$ looks much more involved, it can also be
understood in terms of appropriately defined chiral currents, as will be seen in the next subsection using a self-dual
sigma-model.

If we were considering a canonical transformation with generating function $-{\cal F}$, flipping the sign of ${\cal F}$
in defining equation \eqn{genafunc},
$\tilde{X}_1$ would be replaced by $-\tilde{X}_1$ and Buscher rules would be a variant of equations \eqn{toutaforma},
picking a minus sign in the transformation law of $\tilde{G}_{1i}$ and $\tilde{B}_{1i}$. Then, $\tilde{E}_{1i}$ and
$\tilde{E}_{i1}$ (but not $E_{1i}$ and $E_{i1}$, since $X^1$ remains inert) should flip signs in the corresponding expressions
\eqn{dualcoordinates} and \eqn{tsitsiol} relating $X^1$ with $\tilde{X}_1$ and T-duality would manifest
as $\tilde{{X}}_{1L} = - X^1_L$ and $\tilde{{X}}_{1R} = X^1_R$ instead of $\tilde{{X}}_{1L} = X^1_L$ and
$\tilde{{X}}_{1R} = - X^1_R$ as a matter of convention.

\subsection{T-duality in terms of chiral currents}

According to Ro\v{c}ek and Verlinde,
any pair of dual sigma-models can be represented as quotients of it, showing that T-duality transformations
act as automorphism on the algebra of conformal fields \cite{Rocek:1991ps}. The construction is reminiscent of
modern day double field theory, though here it is taken on the world-sheet, since the master action includes the
Killing coordinate and its dual counterpart
on equal footing. We outline the main steps and results that will be used later in the calculations.

First, let us consider a variant of the two-dimensional non-linear sigma-model \eqn{wsaction}, which is
naturally defined by the action
\begin{equation}\label{wsaction4}
S^{\prime} ={1 \over 2\pi \alpha^{\prime}} \int dz d \bar{z} \left( g_{ij}(X) + b_{ij}(X) \right) \partial X^i \bar{\partial} X^j \ .
\end{equation}
Here, the target space is $(d-1)$-dimensional with coordinates $X^i$ instead of $X^I = (X^1, X^i)$ that appear in the
original action $S$. The metric $g_{ij}$ and the anti-symmetric tensor field $b_{ij}$ are not identical to $G_{ij}$ and
$B_{ij}$ of the original action, by restricting them on the orbits of the Killing vector field $\xi = \partial / \partial X^1$,
but they will be related to them later.

Next, we enlarge the target space by introducing two additional coordinates $X_L^1$ and $X_R^1$, which are a priori unrelated to the
Killing coordinate $X^1$ of $S$, and define a new $(d+1)$-dimensional "parent" sigma-model
\ba
S_{\rm LR} & = & S^{\prime} + {1 \over 2 \pi \alpha^{\prime}} \int dz d \bar{z} \Big[
\partial X_L^1 \bar{\partial} X_L^1 + \partial X_R^1 \bar{\partial} X_R^1 +
2 B(X) \partial X_R^1 \bar{\partial} X_L^1 + \Big. \nonumber\\
& & \Big. ~~~~~~~~ 2 G_i^L (X) \partial X^i \bar{\partial} X_L^1 +
2 G_i^R (X) \bar{\partial} X^i \partial X_R^1 \Big]
\label{wsaction9}
\ea
whose couplings $B(X)$, $G_i^L(X)$ and $G_i^R(X)$ depend only on the $d-1$ coordinates $X^i$. With some additional
ingredients that will be introduced shortly, $X_L^1$ and $X_R^1$ help to reconstruct the coordinate $X^1$
and its dual $\tilde{X}_1$, making T-duality manifest.
The enlarged action \eqn{wsaction9} admits an $U(1)_L \times U(1)_R$ symmetry under $\delta X_L^1 = \alpha_L (z)$ and
$\delta X_R^1 = \alpha_R (\bar{z})$ with corresponding currents
\ba
{\cal J}^1 (z) & = & \partial X_L^1 + B(X) \partial X_R^1 + G_i^L (X) \partial X^i ~, \nonumber\\
\bar {\cal J}^1 (\bar{z}) & = & \bar{\partial} X_R^1 + B(X) \bar{\partial} X_L^1 + G_i^R (X) \bar{\partial} X^i
\label{thecurrentsm}
\ea
that are chirally conserved on the world-sheet, i.e.,
\be
\bar{\partial} {\cal J}^1 (z) = 0 = \partial \bar {\cal J}^1 (\bar{z}) ~.
\ee

Gauging the symmetry $U(1)_L \times U(1)_R$ completes the construction of the master action.
It is achieved with the aid of
a gauge field with components $(A, \bar{A})$ that couple to $X_L^1$, $X_R^1$ and $X^i$ as follows,
\be
S_{\rm gauged}^{(\pm)} = S_{\rm LR} + {1 \over 2 \pi \alpha^{\prime}} \int dz d \bar{z} \Big[A \bar {\cal J}^1 \pm \bar{A} {\cal J}^1 +
{1 \over 2} (1 \pm B(X)) A \bar{A} \Big] ~.
\label{masteract5}
\ee
The different signs correspond to the two possible ways to gauge the symmetry, employing either axial or vector gauging,
$\delta X_L^1 = \alpha$ and $\delta X_R^1 = \mp \alpha$.
The resulting gauged action \eqn{masteract5} is self-dual in that it encompasses
the original non-linear sigma-model $S$ and its T-dual $\tilde{S}$ on equal footing.

Indeed, integrating out the gauge fields,
we obtain the original action $S$ for one gauging and the dual action $\tilde{S}$ for the other gauging.
The resulting $d$-dimensional dual "child" actions are\footnote{We assume that the world-sheet is topologically trivial to avoid
unnecessary complications that arise by the presence of a total derivative term
$\partial X^1 \bar{\partial} \tilde{X}_1 - \partial \tilde{X}_1 \bar{\partial} X^1$ in the Lagrangian, which is omitted.
Of course, higher genera should be employed to establish T-duality as an exact symmetry of theory to all orders in
string perturbation theory, but they are beyond the scope of the present discussion.}
\ba
S & = & S^{\prime} + {1 \over 2 \pi \alpha^{\prime}} \int dz d \bar{z} \Big[{1 + B \over 1 - B} \partial X^1 \bar{\partial} X^1 +
{2 \over 1-B}G_i^R \partial X^1 \bar{\partial} X^i + \Big. \nonumber\\
& & \Big. ~~~~~~~~~~ {2 \over 1-B} G_i^L \bar{\partial} X^1 \partial X^i +
{2 \over 1-B} G_i^L G_j^R \partial X^i \bar{\partial} X^j \Big]
\label{marida1}
\ea
and
\ba
\tilde{S} & = & S^{\prime} + {1 \over 2 \pi \alpha^{\prime}} \int dz d \bar{z} \Big[{1 - B \over 1 + B} \partial \tilde{X}_1
\bar{\partial} \tilde{X}_1 - {2 \over 1+B}G_i^R \partial \tilde{X}_1 \bar{\partial} X^i + \Big. \nonumber\\
& & \Big. ~~~~~~~~~~ {2 \over 1+B} G_i^L \bar{\partial} \tilde{X}_1 \partial X^i -
{2 \over 1+B} G_i^L G_j^R \partial X^i \bar{\partial} X^j \Big] ~.
\label{marida2}
\ea
The identification is exact provided that the Killing coordinates of $S$ and $\tilde{S}$ are chosen to be
\be
X^1 = X_L^1 + X_R^1 ~, ~~~~~~ \tilde{X}_1 = X_L^1 - X_R^1
\label{marida3}
\ee
and they are required to have the same periodicity. Thus, $X_L^1$ and $X_R^1$ reconstruct the Killing coordinate
and its dual, as in the canonical formulation of T-duality. The
couplings appearing in the action $S_{\rm LR}$ should also be adjusted as
\be
G_{11} = {1 + B \over 1 - B} ~, ~~~~~~~ G_{1i} \mp B_{1i} = {2 \over 1-B} G_i^{R,L} (X)
\ee
and
\be
G_{ij} = g_{ij} + {1 - B \over 2} \left(G_{1i} G_{1j} - B_{1i} B_{1j} \right) , ~~~~~
B_{ij} = - b_{ij} - {1 - B \over 2} \left(G_{1i} B_{1j} - B_{1i} G_{1j} \right) ,
\ee
hereby expressing $B (X)$, $G_i^{R,L} (X)$, $g_{ij} (X)$ and $b_{ij} (X)$ in terms of the background fields $G_{IJ}(X)$ and
$B_{IJ}(X)$ of the $d$-dimensional target space associated to the action $S$.

With these identifications in mind, we obtain the T-duality rules for the components of the metric $G_{IJ}$ and the
anti-symmetric tensor field $B_{IJ}$ by comparing the two actions $S$ and $\tilde{S}$,
\ba
& & \tilde{G}_{11} = {1 \over G_{11}} ~, ~~~~ \tilde{G}_{1i} = {B_{1i} \over G_{11}} ~, ~~~~
\tilde{G}_{ij} = G_{ij} - {G_{1i} G_{1j} - B_{1i} B_{1j} \over G_{11}} ~, \nonumber\\
& & \tilde{B}_{1i} = {G_{1i} \over G_{11}} ~, ~~~~
\tilde{B}_{ij} = B_{ij} - {G_{1i} B_{1j} - B_{1i} G_{1j} \over G_{11}} ~,
\label{toutaforma2}
\ea
which are exactly the transformation rules \eqn{toutaforma}.
Looking at the expressions \eqn{marida1}, \eqn{marida2} and \eqn{marida3}, it follows that the
T-duality rules \eqn{toutaforma2} can be neatly summarized as
\be
B \rightarrow - B ~, ~~~~~~ G_i^{L,R} \rightarrow \pm G_i^{L,R} ~, ~~~~~~ X_L^1 \rightarrow X_L^1 ~, ~~~~~~
X_R^1 \rightarrow - X_R^1 ~.
\ee
An immediate consequence is that the chiral currents \eqn{thecurrentsm} transform under duality as
\be
{\cal J}^1 (z) \rightarrow  {\cal J}^1 (z) ~, ~~~~~~~
\bar {\cal J}^1 (\bar{z}) \rightarrow -\bar {\cal J}^1 (\bar{z}) ~ .
\ee

Thus, T-duality leaves the left-moving current ${\cal J}_1 (z)$ invariant and acts by flipping the sign of the
right-moving current $\bar {\cal J}_1 (\bar{z})$, as in the case of free fields. If the enlarged sigma-model
\eqn{wsaction9} is conformally invariant in the quantum regime, the pair of dual sigma-models $S$ and $\tilde{S}$
will also be conformal; in that case, the resulting models are equivalent as quantum field theories. The
duality among sigma-models, as defined above, is a valid classical operation for non-conformal
backgrounds as well, but in those cases the two models are not equivalent as quantum field theories.

The previous discussion generalizes to backgrounds with $n$ commuting isometries, in which case the enlarged action $S_{\rm LR}$
involves $d+n$ target space coordinates, encompassing the Killing coordinates and their dual counterparts in a unified
way. $S_{\rm LR}$ exhibits an $U(1)_L^n \times U(1)_R^n$ chiral symmetry with associated currents ${\cal J}^a (z)$ and
$\bar{\cal J}^a (\bar{z})$ for each $a = 1, ~2, ~\cdots , ~n$. Gauging all symmetries axially or vectorially
results into two sigma-models that are related by a general $O(n, n)$ T-duality transformation.
Further details can be found in the report \cite{Giveon:1994} and references therein.

It is tempting to use the enlarged action $S_{\rm LR}$ for the world-sheet description of double field theory, which
also encompasses the coordinates and their dual counterparts on equal footing in target space. Work in this direction
is in progress.

\subsection{Combining the two approaches}

Let us now explain the different meaning of the Ro\v{c}ek-Verlinde currents in the enlarged parent theory $S_{LR}$ and
in the reduced child theories $S$ and $\tilde{S}$ obtained by quotient. The key point is that $X_L^1$ and $X_R^1$ are
independent fields of the enlarged theory, but they are interrelated by duality in the reduced theories. The form
of the currents in $S$ and $\tilde{S}$ will be obtained by eliminating $\tilde{X}$ in favor of $X$, and vice versa,
through the canonical formulation of T-duality, hereby combining the two approaches.

As explained before, the parent currents \eqn{thecurrentsm} are chiral by the classical equations of motion of the enlarged
two-dimensional non-linear sigma-model \eqn{wsaction9} and they take the following form
\begin{eqnarray}\label{jdef}
{\cal J}^1(z) = \partial X^1_L + {G_{11} - 1 \over G_{11} + 1} ~ \partial X_R^1 + {1 \over G_{11} + 1} ~ E_{1i} \partial X^i\, ,\nonumber\\
\bar {\cal J}^1(\bar z) = \bar \partial X^1_R + {G_{11} - 1 \over G_{11} + 1} ~ \bar \partial X_L^1 + {1 \over G_{11} + 1} ~
E_{i1} \bar \partial X^i \, .
\end{eqnarray}
They can be rewritten in terms of the target space coordinate $X^1 = X^1_L + X^1_R$ and
$\tilde{X}_1 = X^1_L - X^1_R$, assumed to be independent in $S_{LR}$, as follows,
\ba
& & {\cal J}^1(z) = {1 \over G_{11} + 1} \left(G_{11} \partial X^1 + \partial \tilde{X}_1 + E_{1i} \partial X^i
\right) , \nonumber\\
& & \bar {\cal J}^1(\bar z) = {1 \over G_{11} + 1} \left(G_{11} \bar{\partial} X^1 - \bar{\partial} \tilde{X}_1
+ E_{i1} \bar{\partial} X^i \right) .
\ea
Then, it is immediately obvious that ${\cal J}^1(z)$ remains invariant and $\bar {\cal J}^1(\bar z)$ flips sign under
T-duality, since $\tilde{G}_{11} = 1/G_{11}$, $\tilde{E}_{1i} = E_{1i}/G_{11}$ and $\tilde{E}_{i1} = - E_{i1}/G_{11}$.

The coordinates $X_L^1$ and $X_R^1$ are not chiral, in general, but one can
always define chiral coordinates ${\cal  X}^1_L (z)$ and ${\cal  X}^1_R (\bar{z})$ in the enlarged theory $S_{\rm LR}$, letting
\be
{\cal J}^1(z) = \partial{\cal  X}^1_L (z) ~, ~~~~~~ \bar {\cal J}^1(\bar{z}) = \bar \partial{\cal  X}^1_R (\bar{z}) ~.
\ee
Then, T-duality is reformulated as automorphism
\begin{equation}
\tilde{{\cal X}}_{1L} (z) = {\cal X}^1_L (z) \, , ~~~~~~  \tilde{{\cal X}}_{1R} (\bar{z}) = - {\cal X}^1_R (\bar{z}) \, .
\label{simosi}
\end{equation}
Likewise, one can introduce the following combinations of chiral components
\be
{\cal X}^1 = {\cal X}^1_L + {\cal X}^1_R ~, ~~~~~~ \tilde{\cal X}_1 = {\cal X}^1_L - {\cal X}^1_R
\ee
and restate T-duality in direct analogy with free fields, as
\begin{equation}
{\cal X}^1={\cal X}^1_L+{\cal X}^1_R\,\, \longrightarrow\,\,  \tilde{\cal X}_1={\cal X}_L^1-{\cal X}_R^1\, .
\end{equation}
These new fields are non-locally expressed in terms of the target space coordinates $X^I = (X^1, X^i)$ of the original
sigma-model \eqn{wsaction}. The free fields ${\cal X}^1$ and
$\tilde{\cal X}_1$ exist only in the master theory $S_{\rm LR}$ for general backgrounds. Of course, for ordinary toroidal
backgrounds we have ${\cal X}^1 = X^1$ and $\tilde{\cal X}_1 = \tilde{X}_1$, in which case these fields are well defined
in the coset theories as well.

Next, using the non-local relations \eqn{wstdual} and \eqn{ninlaude} among $X^1$ and $\tilde{X}_1$, we can
eliminate one coordinate in favor of the other and express the parent Ro\v{c}ek-Verlinde currents in the original sigma-model frame $S$ as
\be
{\cal J}^1 = {2 \over G_{11} + 1} \left(G_{11} \partial X^1 + G_{1i} \partial X^i \right) , ~~~~~
\bar {\cal J}^1 = {2 \over G_{11} + 1} \left(G_{11} \bar{\partial} X^1 + G_{1i} \bar{\partial} X^i \right)
\label{alvaria2}
\ee
and in the dual sigma-model frame $\tilde{S}$ as
\be
{\cal J}^1 = {2 \over \tilde{G}_{11} + 1} \left(\tilde{G}_{11} \partial \tilde X^1 + \tilde{G}_{1i} \partial\tilde X^i \right) ,
~~~~~ \bar {\cal J}^1 = - {2 \over \tilde{G}_{11} + 1} \left(\tilde{G}_{11} \bar{\partial}\tilde X^1 +
\tilde{G}_{1i} \bar{\partial} \tilde X^i \right) .
\label{cuka}
\ee
It is immediately obvious that these "child" currents are not chiral fields of the original  sigma-model nor its dual, in general.
Thus, as expected by taking the quotient, the currents ${\cal J}^1$ and $\bar {\cal J}^1$ are
not bona fide conformal fields of the coset theory $S$ nor its dual $\tilde{S}$. The physical operators of the coset conformal
field theory $S$ or $\tilde{S}$ should necessarily commute with the BRST charge of the gauged sigma-model \eqn{masteract5} and they
should be defined up to BRST commutators. T-duality acts by automorphisms only in the self-dual theory $S_{LR}$, whereas
after the gauging the currents \eqn{alvaria2} of one reduced theory are mapped to the currents \eqn{cuka} of the other.

Let us illustrate this point by considering the well known example of the $SU(2)$ WZW model and its gauged
variant $SU(2)/U(1)$. In this case, the $SU(2)$ group elements are parametrized canonically as
\be
g = e^{{i \over 2} X_L \sigma_3} e^{{i \over 2} \theta \sigma_1} e^{{i \over 2} X_R \sigma_3}
\ee
and the action of the $SU(2)$ WZW model for the coordinates $\theta$, $X_L$ and $X_R$ is the enlarged action
$S_{LR}$, which is subsequently gauged. The corresponding Ro\v{c}ek-Verlinde currents for the $U(1)_L \times U(1)_R$
symmetry that is gauged are (dropping the index 1 for convenience)
\be
{\cal J} (z) = \partial X_L + {\rm cos}(2 \theta) \partial X_R \, , ~~~~~~
\bar {\cal J} (\bar{z}) = \bar{\partial} X_R + {\rm cos}(2 \theta) \bar{\partial} X_L
\label{cartani}
\ee
and they correspond to the Cartan generators of the $SU(2)_L \times SU(2)_R$ current algebra of the $SU(2)$ WZW model.
Here, we have $B = {\rm cos} (2 \theta)$, accounting for the anti-symmetric tensor field of the $SU(2)$ model in the
$X_L$ and $X_R$ directions, whereas $G_i^{L,R} = 0$. Setting $X= X_L + X_R$, $\tilde{X} = X_L - X_R$ and noting that
$G_{11} = {\rm cot}^2 \theta = 1/\tilde{G}_{11}$, we obtain by gauging the two dual faces of the $SU(2)/U(1)$ WZW
model described by the actions $S$ and $\tilde{S}$ with respective target space metrics
\be
ds^2 = d \theta^2 + {\rm cot}^2 \theta \, dX^2
\ee
and
\be
d\tilde{s}^2 = d \theta^2 + {\rm tan}^2 \theta \, d\tilde{X}^2 .
\ee
There is also a corresponding dilaton, required by conformal invariance, but no anti-symmetric tensor field.

The currents \eqn{cartani} are chiral, i.e., $\bar{\partial} {\cal J}(z) = 0 = \partial \bar {\cal J} (\bar{z})$
by the equations of motion of the $SU(2)$ WZW model, and they can be rewritten as
\be
{\cal J} (z) = {\rm cos}^2 \theta \, \partial X + {\rm sin}^2 \theta \, \partial \tilde{X} \, , ~~~~~~
\bar {\cal J} (z) = {\rm cos}^2 \theta \, \bar{\partial} X - {\rm sin}^2 \theta \, \bar{\partial} \tilde{X} \, ,
\label{ghyut}
\ee
treating $X$ and $\tilde{X}$ as independent fields of the enlarged theory $S_{LR}$. After gauging, $X$ and $\tilde{X}$
are not independent, but they are dual to each other related by $\partial \tilde{X} = {\rm cot}^2 \theta \, \partial X$
and $\bar{\partial} \tilde{X} = - {\rm cot}^2 \theta \, \bar{\partial} X$. Then, in the reduced theory $S$ the two
currents take the form
\be
{\cal J} (z, \bar{z}) = 2 {\rm cos}^2 \theta \, \partial X \, , ~~~~~~~
\bar {\cal J} (z, \bar{z}) = 2 {\rm cos}^2 \theta \, \bar{\partial} X \, ,
\ee
whereas in the reduced theory $\tilde{S}$ they are expressed as
\be
{\cal J} (z, \bar{z}) = 2 {\rm sin}^2 \theta \, \partial \tilde{X} \, , ~~~~~~~
\bar {\cal J} (z, \bar{z}) = - 2 {\rm sin}^2 \theta \, \bar{\partial} \tilde{X} \, .
\ee
It can be easily seen, using the classical equations of motion following from $S$ or $\tilde{S}$, respectively,
that these are no longer chiral after gauging, hence, their dependence on both $z$ and $\bar{z}$ indicated above.
Also, it becomes obvious that the dual ${\cal J}$ of $S$ is the ${\cal J}$ of $\tilde{S}$ and the dual
$\bar {\cal J}$ of $S$ is the $- \bar {\cal J}$ of $\tilde{S}$, unlike the left and right components of the currents
\eqn{ghyut} of the ungauged $SU(2)$ WZW model, which are self-dual and anti-self-dual, respectively.

Similar considerations apply to the toroidal flux vacua of closed string theory that will be considered next in detail.

\section{Parabolic flux vacua of closed stings}
\setcounter{equation}{0}

Let us first consider the case of a 3-torus, $T^3$, with $G_{IJ} = \delta_{IJ}$ (taking all radii equal to $1$
to simplify the presentation), constant dilaton and background $H$-flux given by
the field strength of the anti-symmetric tensor field,
\be
H_{IJK} = \partial_I B_{JK} + \partial_K B_{IJ} + \partial_J B_{KI} ~.
\ee
Then, the original sigma-model action \eqn{wsaction} takes the form
\be
S = {1 \over 2 \pi \alpha^{\prime}} \int dz d \bar{z} \left(\delta_{IJ} + B_{IJ} (X) \right)
\partial X^I \bar{\partial} X^J
\label{linaflu}
\ee
and the classical equations of motion are
\be
\partial \bar{\partial} X^I = {1 \over 2} \, {H^I}_{JK} \partial X^J \bar{\partial} X^K .
\label{mastfiel}
\ee

More generally, we can consider flux backgrounds which are $T^3$ fibrations over a base manifold $M$,
but, here, $M$ is taken to be a point to avoid unnecessary technicalities.

When the $H$-flux is not zero, the theory \eqn{linaflu} is not conformally invariant at the
quantum level. Indeed, to lowest order in $\alpha^{\prime}$, the beta functions of the metric,
anti-symmetric tensor field and dilaton are, respectively,
\ba
& & \beta (G_{IJ}) = -{1 \over 4} {H_I}^{KL} H_{JKL} ~, ~~~~~~
\beta (B_{IJ}) = - {1 \over 2} \partial_K {H^K}_{IJ} ~, \nonumber\\
& & \beta (\Phi) = -{23 \over 6} - {1 \over 24} H_{IJK} H^{IJK} .
\ea
The theory becomes conformal (though subcritical) only when the $H$-flux is constant and very weak,
so that ${\cal O}(H^2)$ terms are negligible. In the following, we choose to work in this dilute flux approximation
to implement T-dualities and denote the corresponding conformal field theory where all operations will be made by $CFT_{\rm H}$.

The defining background, which is called $H$-flux vacuum of closed string theory, exhibits Abelian isometries,
but their number depends on the choice of the anti-symmetric tensor field. $B_{IJ}$ is taken to be linear in the toroidal
coordinates, so that $H_{IJK}$ is constant. A particularly convenient choice is
\be
B_{12} = H \, X^3 \, , ~~~~~~~ B_{23} = 0 = B_{31}
\label{marilia1}
\ee
so that $CFT_{\rm H}$ exhibits two commuting isometries labeled by $a=1, \, 2$, whereas $I = 1, \, 2, \, 3$. Another
convenient choice is
\be
B_{12} = {H \over 2} \, X^3 \, , ~~~~~~  B_{23} = 0 \, , ~~~~~~ B_{31} = {H \over 2} \, X^2 ~,
\ee
which accounts for only one isometry associated to the coordinate $X^1$. Finally, there is yet another
(more symmetric) choice
\be
B_{12} = {H \over 3} \, X^3 \, , ~~~~~~~ B_{23} = {H \over 3} \, X^1 \, , ~~~~~~~ B_{31} = {H \over 3} \, X^2
\ee
in which case none of the isometries is manifest and Buscher's rules, which involve the field $B$ rather than $H$,
are not directly applicable. Yet, the various choices of the 2-form field $B$ are related by gauge transformations
$B \rightarrow B + d \Lambda$ for appropriate choices of $\Lambda$ and the T-dual faces of the model should be
gauge independent.

In the following, we stick to the choice \eqn{marilia1} for $B$, which makes manifest the maximum number of
isometries of the model. The $H$-flux background and its T-dual faces belong in the class of toroidal fibrations and they
exhibit non-trivial monodromies. These mathematical aspects of the model, as well as the description of the
so called $f$-flux, $Q$-flux and $R$-flux backgrounds that arise by successive T-duality transformations are
discussed in detail in Appendices A and B, respectively, together with the monodromies of their currents that
encode non-trivial closed string boundary conditions.

\subsection{T-dual coordinates}

Consider the solution to the classical equations of motion of the $H$-flux model in terms of free fields, which
to linear order in $H$ reads
\be
X^I (z, \bar{z}) = X_{0, L}^I (z) + X_{0, R}^I (\bar{z}) + {1 \over 2} \,
{H^I}_{JK}  X_{0, L}^J (z)  X_{0, R}^K (\bar{z})
\label{loucaki1}
\ee
for all $I=1, \, 2, \, 3$. This expression is independent of the gauge choice for the 2-form field B, which only affects the
form of the dual coordinates. Our task is to find similar expressions for the dual coordinates in terms of free fields for
the choice \eqn{marilia1}.

Recall that the dual coordinate to $X^1$ is defined by the following general relations
\be
\partial \tilde{X}_1 = E_{11} \partial X^1 + E_{i1} \partial X^i , ~~~~~~
\bar{\partial} \tilde{X}_1 = -(E_{11} \bar{\partial} X^1 + E_{1i} \bar{\partial} X^i) \, .
\label{loucaki3}
\ee
For the model at hand we have $E_{11} = 1$ and $E_{1i} = - E_{i1} = B_{1i}$ and so
\be
\partial \tilde{X}_1 = \partial X^1 - H X^3 \partial X^2 , ~~~~~~
\bar{\partial} \tilde{X}_1 = - \bar{\partial} X^1 - H X^3 \bar{\partial} X^2 .
\label{loucaki4}
\ee
Substituting the free field realization \eqn{loucaki1} of the toroidal coordinates, we obtain
\ba
& & \partial \tilde{X}_1 = \partial X_{0, L}^1 - H X_{0, L}^3 \partial X_{0, L}^2
- {H \over 2} \partial (X_{0, L}^2 X_{0, R}^3 + X_{0, L}^3 X_{0, R}^2) \, , \\
& & \bar{\partial} \tilde{X}_1 = - \bar{\partial} X_{0, R}^1 - H X_{0, R}^3 \bar{\partial} X_{0, R}^2
- {H \over 2} \bar{\partial} (X_{0, L}^2 X_{0, R}^3 + X_{0, L}^3 X_{0, R}^2) \, .
\ea
Integration is straightforward and it yields the following result for the dual coordinate
\ba
\tilde{X}_1 (z, \bar{z}) & = & X_{0, L}^1(z) - X_{0, R}^1(\bar{z}) -
{H \over 2} \left(X_{0, L}^2(z) X_{0, R}^3(\bar{z}) + X_{0, L}^3(z) X_{0, R}^2(\bar{z})\right) - \nonumber\\
& & H \int^z X_{0, L}^3(w) \partial X_{0, L}^2(w) d w - H \int^{\bar{z}} X_{0, R}^3(\bar{w})
\bar{\partial} X_{0, R}^2(\bar{w}) d \bar{w} \, ,
\ea
whereas the corresponding expression for the original coordinate $X^1$, following from eq. \eqn{loucaki1}, is
\be
X^1 (z, \bar{z}) = X_{0, L}^1(z) + X_{0, R}^1(\bar{z}) +
{H \over 2} \left(X_{0, L}^2(z) X_{0, R}^3(\bar{z}) - X_{0, L}^3(z) X_{0, R}^2(\bar{z})\right) \, .
\label{loucaki8}
\ee

Next, we define the fields
\be
X_L^1 (z, \bar{z}) = {1 \over 2} (X^1 + \tilde{X}_1) \, , ~~~~~~
X_R^1 (z, \bar{z}) = {1 \over 2} (X^1 - \tilde{X}_1) \,
\ee
in terms of which T-duality reads
\be
X_L^1 (z, \bar{z}) \rightarrow X_L^1 (z, \bar{z}) \, , ~~~~~~
X_R^1 (z, \bar{z}) \rightarrow - X_R^1 (z, \bar{z}) \,.
\ee
Their free field realization follows from above, leading to the expressions
\ba
& & X_L^1 (z, \bar{z}) = X_{0, L}^1 (z) - {H \over 2} X_{0, L}^3(z) X_{0, R}^2(\bar{z})
- {H \over 2} \int^z X_{0, L}^3(w) \partial X_{0, L}^2(w) d w - \nonumber\\
& & ~~~~~~~~~~~~~~~~~~~ {H \over 2}
\int^{\bar{z}} X_{0, R}^3(\bar{w}) \bar{\partial} X_{0, R}^2(\bar{w}) d \bar{w} \, , \\
& & X_R^1 (z, \bar{z}) = X_{0, R}^1 (\bar{z}) + {H \over 2} X_{0, L}^2(z) X_{0, R}^3(\bar{z})
+ {H \over 2} \int^z X_{0, L}^3(w) \partial X_{0, L}^2(w) d w + \nonumber\\
& & ~~~~~~~~~~~~~~~~~~~ {H \over 2}
\int^{\bar{z}} X_{0, R}^3(\bar{w}) \bar{\partial} X_{0, R}^2(\bar{w}) d \bar{w} \, .
\label{loucaki12}
\ea
Thus, the effect of T-duality, going from the $H$-flux to the $f$-flux background, can be neatly expressed in
terms of free fields as follows,
\ba
& & X_{0, L}^1 (z) \rightarrow X_{0, L}^1 (z) \, , \\
& & X_{0, R}^1 (\bar{z}) \rightarrow - X_{0, R}^1 (\bar{z}) - H X_{0, L}^2(z) X_{0, R}^3(\bar{z}) \nonumber\\
& & ~~~~~~~~~~~~~~~~~~ - H \int^z X_{0, L}^3(w) \partial X_{0, L}^2(w) d w -
H \int^{\bar{z}} X_{0, R}^3(\bar{w}) \bar{\partial} X_{0, R}^2(\bar{w}) d \bar{w} \, ,
\ea
leaving $X_{0, L}^i(z)$ and $X_{0, R}^i(\bar{z})$ inert for the remaining toroidal coordinates $i=2, \, 3$.

Let us now prepare to pass to the $Q$-flux model by performing T-duality along the direction $X^2$ of the $f$-flux model.
On the $f$-flux background there is an off-diagonal component
of the metric $G_{12} = fX^3$, identifying $H \equiv f$, whereas all diagonal components are 1 as before. Furthermore,
the 2-form field vanishes.
The dual coordinate to $X^2$ is defined by specializing the defining relations \eqn{loucaki3} to the $f$-flux model whose
coordinates are $\tilde{X}_1, \, X^2, \, X^3$. We have
\be
\partial \tilde{X}_2 = \partial X^2 + H X^3 \partial \tilde{X}_1 , ~~~~~~
\bar{\partial} \tilde{X}_2 = - \bar{\partial} X^2 - H X^3 \bar{\partial} \tilde{X}_1 .
\ee
Employing equation \eqn{loucaki4} we can eliminate $\tilde{X}_1$ and express everything in terms of the coordinates
$X^I$ of the original $H$-flux model. To linear order in $H \equiv f \equiv Q$ the result is
\be
\partial \tilde{X}_2 = \partial X^2 + H X^3 \partial X^1 , ~~~~~~
\bar{\partial} \tilde{X}_2 = - \bar{\partial} X^2 + H X^3 \bar{\partial} X^1 .
\label{loucaki16}
\ee
Note that these are precisely the defining relations of $\tilde{X}_2$ if we were to perform a T-duality
of the $H$-flux model along the $X^2$ direction alone. Thus, the dual coordinates can be invariantly defined
using the $H$-flux model and it does not matter if another duality has been performed first.

According to equation \eqn{loucaki1}, the original coordinate $X^2$ admits the following free field realization,
\be
X^2 (z, \bar{z}) = X_{0, L}^2(z) + X_{0, R}^2(\bar{z}) -
{H \over 2} \left(X_{0, L}^1(z) X_{0, R}^3(\bar{z}) - X_{0, L}^3(z) X_{0, R}^1(\bar{z})\right) \, .
\label{loucaki17}
\ee
Then, comparing \eqn{loucaki4} to \eqn{loucaki16} and \eqn{loucaki8} to \eqn{loucaki17}, we see that the
corresponding expressions are simply related by exchanging the coordinate indices 1 and 2 and flipping the sign of
$H$ at the same time. Implementing these operations to equations (3.17) and \eqn{loucaki12}, we immediately
obtain the following result for the left- and right-components of $X^2$ and $\tilde{X}_2$, respectively,
\ba
& & X_L^2 (z, \bar{z}) = X_{0, L}^2 (z) + {H \over 2} X_{0, L}^3(z) X_{0, R}^1(\bar{z})
+ {H \over 2} \int^z X_{0, L}^3(w) \partial X_{0, L}^1(w) d w + \nonumber\\
& & ~~~~~~~~~~~~~~~~~~~ {H \over 2}
\int^{\bar{z}} X_{0, R}^3(\bar{w}) \bar{\partial} X_{0, R}^1(\bar{w}) d \bar{w} \, , \\
& & X_R^2 (z, \bar{z}) = X_{0, R}^2 (\bar{z}) - {H \over 2} X_{0, L}^1(z) X_{0, R}^3(\bar{z})
- {H \over 2} \int^z X_{0, L}^3(w) \partial X_{0, L}^1(w) d w - \nonumber\\
& & ~~~~~~~~~~~~~~~~~~~ {H \over 2}
\int^{\bar{z}} X_{0, R}^3(\bar{w}) \bar{\partial} X_{0, R}^1(\bar{w}) d \bar{w} \, .
\ea

Finally, combining the above, we see that the effect of T-duality, going from the $H$-flux to the
$Q$-flux background, can be neatly expressed in terms of free fields as
\ba
& & X_{0, L}^1 (z) \rightarrow X_{0, L}^1 (z) - H X_{0, L}^3(z) X_{0, R}^2(\bar{z})
- H \int^{\bar{z}} X_{0, R}^3(\bar{w}) \bar{\partial} X_{0, R}^2(\bar{w}) d \bar{w} \, , \\
& & X_{0, R}^1 (\bar{z}) \rightarrow - X_{0, R}^1 (\bar{z}) - H X_{0, L}^2(z) X_{0, R}^3(\bar{z})
- H \int^z X_{0, L}^3(w) \partial X_{0, L}^2(w) d w
\ea
and
\ba
& & X_{0, L}^2 (z) \rightarrow X_{0, L}^2 (z) + H X_{0, L}^3(z) X_{0, R}^1(\bar{z})
+ H \int^{\bar{z}} X_{0, R}^3(\bar{w}) \bar{\partial} X_{0, R}^1(\bar{w}) d \bar{w} \, , \\
& & X_{0, R}^2 (\bar{z}) \rightarrow - X_{0, R}^2 (\bar{z}) + H X_{0, L}^1(z) X_{0, R}^3(\bar{z})
+ H \int^z X_{0, L}^3(w) \partial X_{0, L}^1(w) d w \, ,
\ea
whereas $X_{0, L}^3(z)$ and $X_{0, R}^3(\bar{z})$ remain inert.

\subsection{Currents and generalized coordinates}

The Ro\v{c}ek-Verlinde approach of gauging the isometries does not require conformality and, as such, it is also
applicable to tori with large $H$-flux. Here, we are limiting ourselves to $CFT_{\rm H}$.
Then, the corresponding currents \eqn{alvaria2} take the simple form
\be
{\cal J}^a = \partial X^a , ~~~~~~~ \bar {\cal J}^a = \bar{\partial} X^a
\ee
in terms of the toroidal coordinates $a=1, \, 2$.
Using the solution \eqn{loucaki1} of the classical equations of motion of the $H$-flux model with very weak constant flux,
the Ro\v{c}ek-Verlinde parent currents take following form in terms of free fields,
\ba\label{gencurrents}
& & {\cal J}^I(z, \bar{z}) = \partial X^I (z, \bar{z}) =
\partial X_{0, L}^I (z) + {1 \over 2} \, {H^I}_{JK} \partial X_{0, L}^J (z) \, X_{0, R}^K (\bar{z}) \, \nonumber\\
& & \bar {\cal J}^I(z, \bar{z}) = \bar{\partial} X^I (z, \bar{z}) =
\bar{\partial} X_{0, R}^I (\bar{z}) + {1 \over 2} \, {H^I}_{JK} X_{0, L}^J (z) \,
\bar{\partial} X_{0, R}^K (\bar{z}) \, .
\label{thisisit}
\ea
Here, the index $I$ is meant to be $a = 1, \, 2$, but for later use we also extend the definition to
encompass ${\cal J}^3(z, \bar{z}) := \partial X^3 (z, \bar{z})$ and $\bar {\cal J}^3(z, \bar{z}) := \bar{\partial} X^3 (z, \bar{z})$,
although the latter have no analogue in the enlarged theory $S_{\rm LR}$.

The currents ${\cal J}^a$ and $\bar {\cal J}^a$ are chiral fields of the enlarged parent theory $S_{\rm LR}$, i.e.,
$\bar{\partial} {\cal J}^a = 0 = \partial \bar {\cal J}^a$, and, thus, they can be readily used to describe the dual
faces of $CFT_{\rm H}$ by the automorphisms ${\cal J}^a \rightarrow {\cal J}^a$ and $\bar {\cal J}^a \rightarrow - \bar {\cal J}^a$.
However, these currents are not chiral in the context of the reduced child theory $CFT_{\rm H}$, since there are obstructions
of order $H$, which arise by the field equations \eqn{mastfiel}. They rather satisfy the non-chiral conservation laws
\be
\bar{\partial} {\cal J}^I - \partial \bar{\cal J}^I = 0 \, ,
\ee
which are immediate and obvious for all $I=1, \, 2, \, 3$. The monodromy properties of those currents are described in Appendix B
for all T-dual faces of the constant flux vacua.

As emphasized before, this difference is attributed to the simple
fact that the Ro\v{c}ek-Verlinde currents do not commute with the BRST charge of the gauged sigma-model associated to the
enlarged theory $S_{\rm LR}$. Our results about non-commutativity and/or non-associativity of the (appropriately defined) coordinates
of the dual faces of $CFT_{\rm H}$ that will be discussed later rely on this subtle point. In fact, the
cohomological interpretation of the resulting algebraic structures in terms of 2- and 3-cocycles can be solely described
in terms of the BRST (i.e., Lie algebra) coboundary operator and it is inherited in the reduced theory by gauging.
Thus, non-commutativity and/or non-associativity arise only in the reduced theory and not in the enlarged theory,
which is manifestly commutative and associative. This interpretation, which will be pursued later in the discussions, 
ties up nicely with the idea that the enlarged theory $S_{\rm LR}$ can be used further to provide the
world-sheet description of double field theory.

The coordinates $X^I$ do not transform in a simple linear way under T-duality. However, we can define {\sl generalized}
coordinates ${\cal  X}^I(z,\bar z)={\cal  X}^I_L(z,\bar z)+{\cal  X}^I_R(z,\bar z)$ that transform nicely under T-duality, as dictated
by eq. (\ref{simosi}). The generalized coordinates are integrals of the  Ro\v{c}ek-Verlinde currents
${\cal J}^I$ and $\bar {\cal J}^I$, given by eq.(\ref{gencurrents}), of the following form,
\begin{equation}
 {\cal  X}^I_L(z,\bar z)=\int^z_a {\cal J}^I(z',\bar z')dz'\, ,\quad  {\cal  X}^I_R(z,\bar z)=
 \int^{\bar z}_{\bar a} \bar {\cal J}^I(z',\bar z')d\bar z'\, .
\end{equation}
Here $a$ and $\bar a$ are some complex integration constants. The expressions above involve integration over the world-sheet coordinate $\sigma$
in the $(\tau,\sigma)$-plane in radial quantization, which is adopted hereafter, and, therefore, they should be defined as
circular integrals in the complex $z$-plane,
keeping $|z|$ and $|\bar z|$ fixed (and, thus, it is also implicitly assumed that $|z|=|a|$ and $|\bar z|=|\bar a|$).
The corresponding fields ${\cal  X}^I$ should be regarded as the proper generalized string coordinates of the $\sigma$-model to first
order in  perturbation theory. The commutation relations of the coordinates ${\cal  X}^I$ and ${\cal  X}^J$, and not of $X^I$ and $X^J$,
and their dual counterparts, will be in focus later.

The different meaning of $({\cal J}^I, \, \bar {\cal J}^I)$ in $CFT_{\rm H}$ and $S_{\rm LR}$ is also reflected
in their operator product expansions. In $CFT_{\rm H}$, we use
the free field 2-point functions
\be
<\partial X_{0, L}^I (z) \partial X_{0, L}^{I^{\prime}} (w)> = {\delta^{I, I^{\prime}} \over 2 (z-w)^2} \, , ~~~~~~
<\bar{\partial} X_{0, R}^I (\bar{z}) \bar{\partial} X_{0, R}^{I^{\prime}} (\bar{w})> =
{\delta^{I, I^{\prime}} \over 2 (\bar{z}-\bar{w})^2}
\ee
to obtain the following expansions, to linear order in $H$,
\ba
& & {\cal J}^I (z, \bar{z}) {\cal J}^{I^{\prime}} (w, \bar{w}) = {\delta^{I, I^{\prime}} \over 2(z-w)^2} + {1 \over 4} \,
{H^{I I^{\prime}}}_K {\bar{z} - \bar{w} \over (z-w)^2} \, \bar{{\cal J}}^K (w, \bar{w}) + \cdots \, , \\
& & \bar{{\cal J}}^I (z, \bar{z}) \bar{{\cal J}}^{I^{\prime}} (w, \bar{w}) = {\delta^{I, I^{\prime}} \over 2 (\bar{z}-\bar{w})^2} +
{1 \over 4} \, {H^{I I^{\prime}}}_K {z - w \over (\bar{z}-\bar{w})^2} \, {\cal J}^K (w, \bar{w}) + \cdots \, , \\
& & {\cal J}^I (z, \bar{z}) \bar{{\cal J}}^{I^{\prime}} (w, \bar{w}) = {1 \over 4} \, {H^{I I^{\prime}}}_K
\left({{\cal J}^K (w, \bar{w}) \over \bar{z} - \bar{w}} - {\bar{{\cal J}}^K (w, \bar{w}) \over z - w} \right) + \cdots \, ,
\ea
where dots denote non-singular terms. There is no holomorphic factorization of the current algebra because of the
mixed terms involving $X_{0, L}^I (z)$ and $X_{0, R}^I (\bar{z})$ in the free field realization \eqn{thisisit}
via the fluxes. In the enlarged theory $S_{\rm LR}$, on the other hand, holomorphic factorization is
guaranteed by construction, leading to the operation product expansion of a $U(1)_L \times U(1)_R$ current algebra
for each generator,
\be
{\cal J}^I (z) {\cal J}^{I^{\prime}} (w) = {\delta^{I, I^{\prime}} \over 2 (z-w)^2} \, , ~~~~~~
\bar{{\cal J}}^I (\bar{z}) \bar{{\cal J}}^{I^{\prime}} (\bar{w}) = {\delta^{I, I^{\prime}} \over 2 (\bar{z}-\bar{w})^2} \, ,
~~~~~~ {\cal J}^I (z) \bar{{\cal J}}^{I^{\prime}} (\bar{w}) = 0 \, ,
\label{maroul5}
\ee
up to non-singular terms that are irrelevant.

Strictly speaking, the index $I$ should be restricted to the Killing directions $1$ and $2$, but their operator
product expansion gives rise to ${\cal J}^3$ and $\bar {\cal J}^3$ in the context of $CFT_{\rm H}$. These additional currents
are not associated to an isometry and they are the ones that obstruct holomorphic factorization. Note, however, that it is not
consistent with the free field realization \eqn{thisisit} to project them out by forming an ideal of the algebra. Their
appearance in the same multiplet with the ordinary Ro\v{c}ek-Verlinde currents should be held responsible for the
emergence of the non-commutative and non-associative structure when T-duality is formally performed in the
third non-Killing direction to yield the non-geometric $R$-flux vacuum of closed string theory.

\section{Canonical T-duality and non-commutativity}

This section is devoted to the derivation of the non-commutative and non-associative relations among the generalized
coordinates appearing in the different T-dual faces of the constant flux background. As such, it contains our main
results, whereas all other material in the paper can be regarded as prerequisite for their derivation.

\subsection{Preliminaries: the case of free fields}

First, we briefly discuss the commutation relations among free fields, using the formalism of conformal field theory, as warm up,
while preparing the ground for the derivation of the
commutation relations among the generalized coordinates of the flux backgrounds by similar techniques.
Recall that the free fields $X_0^I(z,\bar z)=X_{0,L}^I(z)+X_{0,R}^I(\bar z)$ possesses the following mode expansion
\begin{eqnarray}
X_{0,L}^I(z)&=&x_{0,L}^I-ip_L^I\log z+i\sum_{n\neq 0}{1\over n}\alpha_n^Iz^{-n}
\, ,\nonumber\\
X_{0,R}^I(\bar z)&=&x_{0,R}^I-ip_R^I\log \bar z+i\sum_{n\neq 0}{1\over n}\bar\alpha_n^I\bar z^{-n}\, ,
\end{eqnarray}
introducing complex coordinates as $z=e^{\tau-i\sigma}$ and $\bar z =e^{\tau+i\sigma}$.
All commutators are derived from the free field propagators of the left- and right-moving fields, which are taken to be
\begin{eqnarray}\label{freefieldcorAA}
 \langle X_{0,L}^I(z)         X_{0,L}^J(w)\rangle&=&{1\over 2}\delta^{IJ}\log(z-w)\, ,\nonumber\\
 \langle X_{0,R}^I(\bar z)         X_{0,R}^J(\bar w)\rangle&=&{1\over 2}\delta^{IJ}\log(\bar z-\bar w)\, ,\nonumber\\
 \langle X_{0,L}^I( z)         X_{0,R}^J(\bar w)\rangle&=&0\,
\end{eqnarray}
with the appropriate normalization of the string tension.

Let us compute the equal time commutator of the left-moving coordinate $X_{0,L}^I(z)$ with its derivative $\partial_w X_{0,L}^J(w)$.
Time ordering becomes radial ordering in the complex $z$-plane, and, thus, the equal time commutator
$\lbrack  X_{0,L}^I(\tau,\sigma), ~ \partial_w X_{0,L}^J(\tau,\sigma')\rbrack$ is defined as follows, taking into account the
$\delta$-prescription,
\begin{eqnarray}
&{~}&\lbrack  X_{0,L}^I(\tau,\sigma), ~ \partial_w X_{0,L}^J(\tau,\sigma')\rbrack=\lbrack  X_{0,L}^I(z), ~
\partial_w X_{0,L}^J(w)\rbrack_{|z|=|w|}=
\nonumber \\ &{~}&\hskip0.1cm = \lim_{\delta\rightarrow 0_+}\biggl(\bigl(X_{0,L}^I(z)\partial_wX_{0,L}^J(w)\bigr)_{|z|=|w|+\delta}-\bigl(\partial_wX_{0,L}^J(w)
X_{0,L}^I(z)\bigr)_{|z|=|w|-\delta}\biggr) .
\end{eqnarray}
Using  the two-point function (\ref{freefieldcorAA}),  we obtain by contraction that
\begin{eqnarray}
\lbrack  X_{0,L}^I(z), ~ \partial_wX_{0,L}^J(w)\rbrack_{|z|=|w|}=
-{1\over 2} \delta^{IJ}\lim_{\delta\rightarrow 0_+}\biggl({1\over z-w}|_{|z|=|w|+\delta}-{1\over z-w}|_{|z|=|w|-\delta}
\biggr)
\end{eqnarray}
and so, choosing $z=e^{\pm\delta-i\sigma}$, $w=e^{-i\sigma'}$, i.e., $\tau=\pm\delta$, $\tau'=0$ and $\varphi=\sigma-\sigma'$,
the above commutator works out to be
\begin{eqnarray}
\lbrack  X_{0,L}^I(z), ~ \partial_wX_{0,L}^J(w)\rbrack_{|z|=|w|}&=&
 -{1\over 2}\delta^{IJ} \lim_{\delta\rightarrow 0_+}\biggl({e^{-\delta +i\sigma}\over 1-e^{-\delta+i\varphi}}+  {e^{i\sigma'}\over 1-e^{-\delta-i\varphi} }  \biggr)\nonumber\\
&=&-{1\over 2}\delta^{IJ}\biggl(e^{i\sigma}\sum_{m=0}^\infty e^{im\varphi}+e^{i\sigma'}\sum_{m=0}^\infty e^{-im\varphi}\biggr)\nonumber\\
&=&
-\pi\delta^{IJ} e^{i\sigma'}\delta(\varphi)=i\pi\delta^{IJ}\delta(z-w)\, .
\end{eqnarray}

Next, we compute the equal time commutator of $X_{0,L}^I(z)$ with  $X_{0,L}^J(w)$, representing $X_{0,L}^J(w)$ by the integral
\begin{equation}
X_{0,L}^J(w)=\int^w d\tilde w~\partial_{\tilde w} X_{0,L}^J(\tilde w)
\end{equation}
in view of the generalizations that will be considered later.
The integral goes over some path in the complex $w$-plane which is not a closed contour. The computation of the equal time
commutator dictates choosing a path of constant time, thus integrating along the circular direction in the complex plane by rewriting
\begin{equation}
X_{0,L}^J(w)=-i\int^{\sigma'} d\tilde \sigma'e^{-i\tilde \sigma'}\partial_{\tilde w} X_{0,L}^J(\tilde w) \, .
\end{equation}
Then, the commutator
 $\lbrack  X_{0,L}^I(\tau,\sigma), ~ X_{0,L}^J(\tau,\sigma')\rbrack$ is computed in steps as follows,
\begin{eqnarray}\label{comfree1}
\lbrack X_{0,L}^I(\tau,\sigma), ~ X_{0,L}^J(\tau,\sigma')\rbrack&=&\lbrack  X_{0,L}^I(z), ~ X_{0,L}^J(w)\rbrack_{|z|=|w|}\nonumber\\
&=&\lbrack  X_{0,L}^I(z),     ~  \int^w d\tilde w~\partial_{\tilde w} X_{0,L}^J(\tilde w)       \rbrack_{|z|=|w|}\nonumber\\
&=&-i\int^{\sigma'} d\tilde \sigma'e^{-i\tilde \sigma'}\lbrack  X_{0,L}^I(z), ~ \partial_{\tilde w}X_{0,L}^J(\tilde w)\rbrack_{|z|=|\tilde w|}\nonumber\\
&=&i\pi\delta^{IJ}\int^\varphi d\tilde\varphi~\delta(\tilde\varphi)
={i\pi\over 2}\delta^{IJ}~\epsilon(\varphi)\, ,
\end{eqnarray}
using also the identity (C.2) found in Appendix C. The same result is obtained directly from the
logarithmic expression (\ref{freefieldcorAA}) for the propagator, without making use of the integral formula above.

Using the same techniques, one obtains similar expressions for the commutators for the right-moving free fields
\begin{eqnarray}\label{comfree2}
\lbrack X_{0,R}^I(\tau,\sigma), ~ X_{0,R}^J(\tau,\sigma')\rbrack=\lbrack  X_{0,R}^I(\bar z), ~ X_{0,R}^J(\bar w)\rbrack_{|\bar z|=|\bar w|}
=-{i\pi\over 2}\delta^{IJ}~\epsilon(\varphi)\, .
\end{eqnarray}

Finally, combining the expressions (\ref{comfree1}) and (\ref{comfree2}), the equal time commutators
among $X_0^I(z,\bar z)=X_{0,L}^I(z)+X_{0,R}^I(\bar z)$ and the dual coordinates $\tilde X_0^J(z,\bar z)=X_{0,L}^J(z)-X_{0,R}^J(\bar z)$
work out to be
\begin{eqnarray}
\lbrack X_{0}^I(\tau,\sigma), ~ X_{0}^J(\tau,\sigma')\rbrack&=&   \lbrack  X_{0,L}^I(z), ~ X_{0,L}^J(w)\rbrack_{|z|=|w|}+
\lbrack  X_{0,R}^I(\bar z), ~ X_{0,R}^J(\bar w)\rbrack_{|\bar z|=|\bar w|}=0\, ,
\nonumber \\
\lbrack \tilde X_{0}^I(\tau,\sigma), ~ \tilde X_{0}^J(\tau,\sigma')\rbrack&=&   \lbrack  X_{0,L}^I(z), ~ X_{0,L}^J(w)\rbrack_{|z|=|w|}+
\lbrack  X_{0,R}^I(\bar z), ~ X_{0,R}^J(\bar w)\rbrack_{|\bar z|=|\bar w|}=0\, ,\nonumber\\
\lbrack X_{0}^I(\tau,\sigma), ~ \tilde X_{0}^J(\tau,\sigma')\rbrack&=&   \lbrack  X_{0,L}^I(z), ~ X_{0,L}(w)^J\rbrack_{|z|=|w|}-
\lbrack  X_{0,R}^I(\bar z), ~ X_{0,R}^J(\bar w)\rbrack_{|\bar z|=|\bar w|}
\nonumber\\
&=&i\pi\delta^{IJ}~\epsilon(\varphi)\, ,
\end{eqnarray}
as expected. Since $\epsilon(0)=0$, we have $\lbrack X_{0}^I(\tau,\sigma), ~ \tilde X_{0}^J(\tau,\sigma')\rbrack=0$ for $\sigma=\sigma'$.

\subsection{Non-commutativity for interacting fields with constant fluxes}

We are now in position to extract the general structure of the commutation relations among the closed string coordinates
for toroidal backgrounds with constant flux by conformal field theory techniques based on the formalism introduced
in the previous sections.
Analogous calculations were performed earlier, first  in ref. \cite{Lust:2010iy} and then in \cite{Andriot:2012vb},
using the world-sheet operator formalism of parabolic models (but see also \cite{Blair:2014kla} for the computation of
Dirac brackets among the coordinates), and they were subsequently
extended to backgrounds with elliptic monodromies in ref. \cite{Condeescu:2012sp}.
As will be seen here, the CFT treatment is more efficient, making the origin of non-commutativity more transparent compared
to the previous approaches. The main advantage of the CFT formalism is that T-duality acts in a simple
way, flipping the sign of the right-moving generalized coordinates.

The $\sigma$-model coordinates $X^I(z,\bar z)$ are not anymore free fields, but they interact with each other in the presence of fluxes.
Since the left- and right-moving coordinates $X_L^I(z,\bar z)$ and $X_R^I(z,\bar z)$ are not directly related to currents
that transform linearly under duality transformations, we were led to consider the {\sl generalized}
coordinates ${\cal  X}^I$, as discussed in the previous section, which are the integrals of the
Ro\v{c}ek-Verlinde currents ${\cal J}_L^I(z,\bar z)$ and $\bar {\cal J}_R^I(z,\bar z)$. We have
\begin{equation}
{\cal  X}^I(z,\bar z)={\cal  X}^I_L(z,\bar z)+{\cal  X}^I_R(z,\bar z)\, ,
\end{equation}
letting
\begin{equation}
 {\cal  X}^I_L(z,\bar z)=\int^z_a {\cal J}_L^I(z',\bar z')dz'\, ,\quad  {\cal  X}^I_R(z,\bar z)=\int^{\bar z}_{\bar a} \bar {\cal J}_R^I(z',\bar z')d\bar z'\, .
\end{equation}
As explained earlier, T-duality in the $I$-direction acts on the currents ${\cal J}_L^I(z,\bar z)$ and $\bar {\cal J}_R^I(z,\bar z)$  as well as on
the corresponding generalized coordinates ${\cal  X}^I_L(z,\bar z)$ and ${\cal  X}^I_R(z,\bar z)$ in a simple linear way,
\begin{eqnarray}\label{tdualcal}
& &\quad {\cal  J}^I_L(z,\bar z)\longrightarrow  {\cal  J}^I_L(z,\bar z)\, ,\quad\bar {\cal  J}^I_R(z,\bar z)\longrightarrow  -
\bar{\cal  J}^I_R(z,\bar z)\,,\nonumber\\
& &\quad {\cal  X}^I_L(z,\bar z)\longrightarrow  {\cal  X}^I_L(z,\bar z)\, ,\quad {\cal  X}^I_R(z,\bar z)\longrightarrow  -
{\cal  X}^I_R(z,\bar z)\,.
 \end{eqnarray}
This does not mean that the T-dual faces of the interacting theory are self-dual, but it is a statement that T-duality maps the
generalized coordinates of one flux background to the generalized coordinates of another and vice-versa.

It follows from the definition of the coordinates ${\cal  X}^I_L(z,\bar z)$ as integrals of the non-chiral currents
${\cal J}_L^I(z,\bar z)$ that such integrals are not
anymore path-independent, but, in fact, they depend on the chosen path in the complex plane;
the same is also true for the corresponding right-moving objects.
Hence, it is necessary to specify the path of integration in order to assign a well defined meaning to the generalized
coordinates ${\cal  X}^I_L$ and ${\cal  X}^I_R$. Here, we apply
the radial quantization for the computation of the commutators by choosing circular paths around the
origin of the complex plane, keeping $|z|$ and $|\bar z|$ fixed in the integrals, as in the case of free fields.
Then, with this prescription in mind, which we adopt in the following, the generalized coordinates are defined as follows,
\begin{equation}
 {\cal  X}^I_L(z,\bar z)=-i\int^\sigma d\tilde \sigma~ e^{-i\tilde \sigma}{\cal J}_L^I(z,\bar  z)\, ,\quad
 {\cal  X}^I_R(z,\bar z)=i\int^{\sigma} d\tilde \sigma ~e^{i\tilde \sigma}
 \bar {\cal J}_R^I(z,\bar z)\, ,
\end{equation}
letting $z=e^{\tau -i\sigma}$ and $\bar z=e^{\tau +i\sigma}$. It follows immediately that the non-holomorphic terms of
${\cal J}_L^I(z,\bar  z)$, which depend on $\bar z$, also contribute to the integral over $z$; similar remarks apply to
the holomorphic part of the currents $\bar {\cal J}_R^I(z,\bar z)$.

Let us now consider the equal time commutators among the generalized coordinates ${\cal  X}^I(z,\bar z)$ and ${\cal  X}^J(w,\bar w)$,
\begin{eqnarray}
&{~}&\lbrack  {\cal X}^{I}(\tau,\sigma), ~ {\cal X}^{J}(\tau,\sigma')\rbrack=\lbrack  {\cal X}^{I}(z,\bar z), ~
{\cal X}^{J}(w,\bar w)\rbrack_{|z|=|w|}=
\nonumber \\ &{~}&\hskip0.5cm \lim_{\delta\rightarrow 0_+}\biggl(\bigl({\cal X}^{I}(z,\bar z)
{\cal X}^{J}(w,\bar w)\bigr)_{|z|=|w|+\delta}-\bigl({\cal X}^{J}(w,\bar w)
{\cal X}^{I}(z,\bar z)\bigr)_{|z|=|w|-\delta}\biggr) .
\end{eqnarray}
It is convenient to split these commutators in four different pieces, namely
\begin{eqnarray}
\Theta_{LL}^{IJ}(\tau,\sigma,\sigma')&:=&\lbrack {\cal  X}^I_L(z,\bar z), ~ {\cal  X}^J_L(w,\bar w)\rbrack_{|z|=|w|}\, ,
\nonumber \\
\Theta_{RR}^{IJ}(\tau,\sigma,\sigma')&:=&\lbrack {\cal  X}^I_R(z,\bar z), ~ {\cal  X}^J_R(w,\bar w)\rbrack_{|z|=|w|}\, ,
\nonumber \\
\Theta_{LR}^{IJ}(\tau,\sigma,\sigma')&:=&\lbrack {\cal  X}^I_L(z,\bar z), ~ {\cal  X}^J_R(w,\bar w)\rbrack_{|z|=|w|}\, ,
\nonumber \\
\Theta_{RL}^{IJ}(\tau,\sigma,\sigma')&:=&\lbrack {\cal  X}^I_R(z,\bar z), ~ {\cal  X}^J_L(w,\bar w)\rbrack_{|z|=|w|}\, ,
\end{eqnarray}
letting $z=e^{\tau-i\sigma}$ and $w=e^{\tau'-i\sigma'}$. Note that apart from the left- and right-moving commutators
$\Theta_{LL}^{IJ}$ and $\Theta_{RR}^{IJ}$, which are in general non-vanishing, the cross terms
$\Theta_{LR}^{IJ}$ and $\Theta_{RL}^{IJ}$ can also be non-zero, because
the operator products of ${\cal J}_L^I(z,\bar z)$ with ${\cal J}^J_R(z,\bar z)$ (and, likewise, of $\bar {\cal J}_R^I(z,\bar z)$
with ${\cal J}^J_L(z,\bar z)$) contain non-vanishing poles that contribute to their expressions.
Adding the left- and right-moving pieces together, we obtain the following general structure for the commutation relations
between ${\cal  X}^I={\cal  X}_L^I+{\cal  X}_R^I$ and ${\cal  X}^J={\cal  X}_L^J+{\cal  X}_R^J$,
\begin{eqnarray}\label{Thetasum1}
\Theta^{IJ}=\lbrack {\cal  X}^I, ~ {\cal  X}^J\rbrack= \Theta_{LL}^{IJ}+\Theta_{RR}^{IJ}+\Theta_{LR}^{IJ}+\Theta_{RL}^{IJ}\, .
\end{eqnarray}

As emphasized before, ${\cal  X}_{L,R}^I(z,\bar z)$ are the appropriate generalized string coordinates that transform
nicely under T-duality. Then, starting from a given flux background and dualizing in the $I$-direction, the four
commutators pieces transform, following (\ref{tdualcal}), as
\begin{equation}
\Theta_{LL}^{IJ}\rightarrow \Theta_{LL}^{IJ}\, , \quad \Theta_{RR}^{IJ}\rightarrow -\Theta_{RR}^{IJ}\, ,\quad\Theta_{LR}^{IJ}\rightarrow \Theta_{LR}^{IJ}\, ,\quad\Theta_{RL}^{IJ}\rightarrow -\Theta_{RL}^{ij}
\end{equation}
and, as a result, the total commutators in the dual background take the simple form
\begin{eqnarray}\label{Thetasum1}
\Theta^{IJ}_{\rm dual}=\lbrack \tilde{\cal  X}^I,{\cal  X}^J\rbrack= \Theta_{LL}^{IJ}-\Theta_{RR}^{IJ}+\Theta_{LR}^{IJ}-\Theta_{RL}^{IJ}\, .
\end{eqnarray}

Thus, starting from the $H$-flux background and applying two consecutive T-duality transformations in the 1- and 2-directions, we
obtain in steps:

\vskip0.3cm
\noindent (i) {\underline {Geometric $H$-flux background:}}

\vskip0.2cm

\begin{equation}\label{parabolicH}
\Theta^{12}_H=\Theta^{12}_{LL}+\Theta^{12}_{RR}+\Theta^{12}_{LR}+\Theta^{12}_{RL}\, .
\end{equation}

\vskip0.3cm
\noindent (ii) {\underline {Geometric $f$-flux background:}} after a T-duality in the 1-direction, we obtain
\vskip0.2cm

\begin{equation}\label{parabolicf}
\Theta^{12}_f=\Theta^{12}_{LL}-\Theta^{12}_{RR}+\Theta^{12}_{LR}-\Theta^{12}_{RL}\, .
\end{equation}

\vskip0.3cm
\noindent (iii) {\underline {Non-geometric $Q$-flux background:}} another T-duality in the 2-direction yields

\vskip0.2cm

\begin{equation}\label{parabolicQ}
\Theta^{12}_Q=\Theta^{12}_{LL}+\Theta^{12}_{RR}-\Theta^{12}_{LR}-\Theta^{12}_{RL}\, .
\end{equation}

As will be seen next, some of these commutators are zero and others are non-vanishing, leading to non-commutativity
in the corresponding flux vacua.
The same method leads to non-associativity by making another step in the 3-direction to reach the non-geometric $R$-flux background.

\subsection{Case by case computations}

We will compute the individual pieces of the commutators among the generalized coordinates for all T-dual faces of the
parabolic flux model, case by case, using the technical material contained in Appendix C.

\subsubsection{Commutators of the $H$-flux background}

Let us now apply this formalism to extract the commutators of the $H$-flux background.
In the following, we compute the off-diagonal commutators between the generalized coordinates
in the 1- and 2-directions of the $H$-flux background to linear order in the fluxes.
The diagonal commutators between the generalized coordinates and their dual counterparts in the same directions are identical
to those found for the free fields in the previous subsection.

The left- and right-moving currents of the $H$-flux background in the 1- and 2-directions take the following form,
according to eq. (\ref{gencurrents}),
\begin{eqnarray}\label{q1all}
  {\cal J}^1_L (z,\bar z)&=&\partial X_{0,L}^1(z)+
   \frac{1}{2}H(X_{0,R}^3(\bar z)\, \partial X_{0,L}^2(z)-X_{0,R}^2(\bar z)\partial X_{0,L}^3(z))  \;,\nonumber\\
    {\cal J}^2_L(z,\bar z) & =&\partial X_{0,L}^2(z)-
   \frac{1}{2}H(X_{0,R}^3(\bar z)\, \partial X_{0,L}^1(z)-X_{0,R}^1(\bar z)\partial X_{0,L}^3(z))  \,,\nonumber\\
     \bar {\cal J}^1_R(z,\bar z) &=&\bar\partial X_{0,R}^1(\bar z)+
   \frac{1}{2}H(X_{0,L}^2(z)\, \bar\partial X_{0,R}^3(\bar z)-X_{0,L}^3(z)\bar\partial X_{0,R}^2(\bar z))  \;,\nonumber\\
    \bar {\cal J}^2_R (z,\bar z)&=&\bar\partial X_{0,R}^2(\bar z)-
   \frac{1}{2}H(X_{0,L}^1(z)\, \bar\partial X_{0,R}^3(\bar z)-X_{0,L}^3(z)\bar\partial X_{0,R}^1(\bar z))  \;.
   \end{eqnarray}

From these currents we obtain the following {\sl generalized} coordinates in the 1- and 2- directions of the three-torus,
\begin{eqnarray}
  \label{def_coor_04}
 {\cal  X}^1_L(z,\bar z)&=& X^1_{0,L}(z)+
  {\textstyle \frac{1}{2}}\hspace{0.5pt}H \, \int^zdz'\,
  (X_{0,R}^3(\bar z')\, \partial' X_{0,L}^2(z')-X_{0,R}^2(\bar z')\partial' X_{0,L}^3(z')) \, ,
  \nonumber\\
   {\cal  X}^1_R(z,\bar z)&= &X^1_{0,R}(\bar z)+
  {\textstyle \frac{1}{2}}\hspace{0.5pt}H\, \int^{\bar z}d\bar z'\,
  (X_{0,L}^2(z')\, \bar\partial' X_{0,R}^3(\bar z')-X_{0,L}^3(z')\bar\partial' X_{0,R}^2(\bar z'))
 \;,
 \nonumber\\
 {\cal  X}^2_L(z,\bar z)&=& X^2_{0,L}(z)-
  {\textstyle \frac{1}{2}}\hspace{0.5pt}H \, \int^zdz'\,
  (X_{0,R}^3(\bar z')\, \partial' X_{0,L}^1(z')-X_{0,R}^1(\bar z')\partial' X_{0,L}^3(z')) \, ,
  \nonumber\\
   {\cal  X}^2_R(z,\bar z)&= &X^2_{0,R}(\bar z)-
  {\textstyle \frac{1}{2}}\hspace{0.5pt}H\, \int^{\bar z}d\bar z'\,
  (X_{0,L}^1(z')\, \bar\partial' X_{0,R}^3(\bar z')-X_{0,L}^3(z')\bar\partial' X_{0,R}^1(\bar z'))
  .
 \end{eqnarray}

We start with the equal time commutator piece $\Theta_{LL}^{12}$ to linear order in $H$, which is
\begin{eqnarray}\label{thetaLL1}
\Theta_{LL}^{12}(z,w,\bar z,\bar w)&=&\lbrack {\cal  X}^1_{L}(z,\bar z), ~ {\cal  X}^2_L(w,\bar w)\rbrack\nonumber\\
&=&-{1\over 2}H\biggl\lbrack X_{0,L}^1(z), ~ \int^w \partial' X^{1}_{0,L}(z')    {X}^3_{0,R}(\bar z')  dz'\biggr\rbrack\nonumber\\
&-&{1\over 2}H\biggl\lbrack X_{0,L}^2(w), ~ \int^z \partial' X^{2}_{0,L}(z')    {X}^3_{0,R}(\bar z')  dz'\biggr\rbrack\, .
\end{eqnarray}
Its computation proceeds by taking single contractions of the operators that appear in the two terms above - there are no
higher contractions - which effectively means that
the commutators correspond to classical Poisson (respectively Dirac brackets) prior to quantization. The following
contractions contribute to $\Theta_{LL}^{12}$,
\begin{eqnarray}\label{thetaLL2}
\Theta_{LL}^{12}(z,w,\bar z,\bar w)
&=&- \Biggl\lbrack{1\over 2}H\int^w\biggl( \langle  X_{0,L}^1(z) \partial' X^{1}_{0,L}(z')\rangle     {X}^3_{0,R}(\bar z')  \biggr)dz'\nonumber\\
&+&{1\over 2}H \int^z\biggl(
\langle  X_{0,L}^2(w) \partial' X^{2}_{0,L}(z')\rangle     {X}^3_{0,R}(\bar z') \biggr)dz'\Biggr\rbrack_{|z|=|w|+\delta}\nonumber\\
&-&{\rm the ~other~ half ~of ~the ~commutator}
\end{eqnarray}

It can be seen that the 1-point function of the operator  ${X}^3_{0,R}(\bar z')$ remains in the commutator after the contraction.
This looks rather unusual and one may naively expect that the net result for the commutator is zero. Note, however, that we are computing
these expressions in a sector of the string Hilbert space where the left- and right-moving momenta $p_L^3$ and $p_R^3$ are not zero.
For vanishing oscillator numbers, the string level matching constraint implies that
\begin{equation}
p_L^3 =\pm p_R^3\, .
\end{equation}
Here, we concentrate on states with non-trivial winding number $\tilde p^3$, having
\begin{equation}
\quad \tilde p^3=p_L^3-p_R^3 \quad{\rm and}\quad p_L^3=-p_R^3\,.
\end{equation}
This, in turn, implies that the momentum operators $\hat p_{L,R}^3$ possess non-vanishing matrix elements in the winding states,
\begin{equation}
\langle \tilde p^3|\hat p_{L}^3|\tilde p^3\rangle={\tilde p^3\over 2}\, ,
\quad\langle \tilde p^3|\hat p_{R}^3|\tilde p^3\rangle=-{\tilde p^3\over 2}\, .
\end{equation}

The existence of winding states is closely related to the non-trivial monodromy properties of the flux backgrounds, which are
outlined in Appendices A and B.
Specifically, the monodromy induces non-trivial transformations of the coordinates $X^1$, $X^2$ and of
the dual coordinates $\tilde X_1$, $\tilde X_2$ when going around the third circular direction by making the shift
\begin{equation}
X^3\rightarrow X^3+2\pi
\end{equation}
and selects new closed string boundary conditions. In sectors with non-vanishing winding number $\tilde p^3$,
the shift $\sigma\rightarrow \sigma+2\pi$ acts on $X^3$ in the standard way,
\begin{equation}
X^3(e^{-2i\pi }z, ~ e^{2i\pi}\bar z)=X^3(z,\bar z)-2\pi \tilde p^3\, ,
\end{equation}
whereas, for non-vanishing fluxes, the currents ${\cal J}_{L,R}^I$, with $I=1, \, 2$, undergo non-trivial transformation
given in terms a flux dependent monodromy matrix ${\cal M}_{IJ}$, as
\begin{equation}
{\cal J}_{L,R}^I(e^{-2i\pi }z, ~ e^{2i\pi}\bar z)={\cal M}_{IJ}~{\cal J}_{L,R}^J(z,\bar z)\, .
\end{equation}

The upshot of this discussion is that the commutators can be non-zero in string sectors with non-vanishing winding number.
The physical origin of this behavior should be attributed to the modified closed string boundary conditions, which reflect the simple
fact that non-local winding strings can capture the non-trivial global properties of the flux background, thus giving rise to
non-commutativity among the generalized coordinates. We shall examine how this is precisely realized in all T-dual faces of
the parabolic flux model. With these explanations in mind, let us now continue with the evaluation of the commutator pieces in the
$H$-flux background and then use the results to extract the total commutators in all other T-dual faces of the model.

Using the mode expansion of free fields, the 1-point function of ${X}^3_{0,R}(\bar z')$, which is given by (\ref{thetaLL2}),
turns out to be
\begin{equation}
\langle \tilde p^3|X^3_{0,R}(\bar z')|\tilde p^3\rangle=i {\tilde p^3\over 2}~\log \bar z'
\end{equation}
in sectors with non-trivial winding number. Furthermore, using the 2-point functions
\begin{equation}
 \langle  X_{0,L}^1(z) \partial' X^{1}_{0,L}(z')\rangle=-{1\over 2(z-z')}\, ,\quad\langle  X_{0,L}^2(w) \partial' X^{2}_{0,L}(z')\rangle =-{1\over 2(w-z')}\, ,
 \end{equation}
we arrive at the following intermediate result for $\Theta_{LL}^{12}$,
\begin{eqnarray}\label{thetaLL22}
\Theta_{LL}^{12}(z,w,\bar z,\bar w)
&=&{i\over 8}\tilde p^3H\Biggl\lbrack\int^w
\Biggl(
{\log \bar z'\over z-z'}     \Biggr)
dz'
+\int^z\Biggl(
{\log \bar z'\over w-z'}
\Biggr)dz'\Biggr\rbrack_{|z|=|w|+\delta}\nonumber\\
&-&{\rm the ~other~ half ~of ~the ~commutator}
\, .
\end{eqnarray}
These integrals are evaluated in Appendix C (see, in particular,  eq. (\ref{typeiii})). Hence, we obtain
\begin{eqnarray}\label{2.29}
\Theta_{LL}^{12}(z,w,\bar z,\bar w)=
-{i\pi\over 4}H\tilde p^3
\biggl(
{\varphi^2\over2 \pi}-{1\over \pi}
 {\cal L}i_2(e^{i\varphi})\biggr)\, .
\end{eqnarray}

Setting $\varphi=0$, it yields
\begin{eqnarray}
\Theta_{LL}^{12}(\tau, \sigma)=
iH\tilde p^3
{\pi^2\over 12}\, .
\end{eqnarray}

Likewise, we obtain the following result for $\Theta_{RR}$,
\begin{equation}\label{2.30}
\Theta_{RR}^{12}(z,w,\bar z,\bar w)=
-{i\pi\over 4}H\tilde p^3
\biggl(
{\varphi^2\over2 \pi}-{1\over \pi}
 {\cal L}i_2(e^{i\varphi})\biggr) ,
\end{equation}
which for $\varphi = 0$ yields
\begin{equation}
\Theta_{RR}^{12} (\tau, \sigma) = iH\tilde p^3
{\pi^2\over 12}\, .
\end{equation}

Next, we compute the commutator piece $\Theta_{LR}$, which is given by the expression
\begin{eqnarray}
\Theta_{LR}^{12}(z,\bar z,w,\bar w)&=&\lbrack {\cal  X}^1_L(z,\bar z), ~ {\cal  X}^2_R(w,\bar w)\rbrack\nonumber\\
&=&-{1\over 2}H\biggl\lbrack X_{0,L}^1(z), ~ \int^{\bar w} X^{1}_{0,L}(z') \bar\partial' X^{3}_{0,R}(\bar z') d\bar z'\biggr\rbrack\nonumber\\
&+&{1\over 2}H\biggl\lbrack X_{0,R}^2(\bar w), ~ \int^z \partial' X^{3}_{0,L}(z')    {X}^2_{0,R}(\bar z')  dz'\biggr\rbrack\, .
\end{eqnarray}
Taking the single contractions, we find
\begin{eqnarray}
\Theta_{LR}^{12}(z,w,\bar z,\bar w)
&=&\Biggl\lbrack-{1\over 2}H\int^{\bar w} \langle X_{0,L}^1(z)  X^{1}_{0,L}(z')\rangle
\bar \partial' X^{3}_{0,R}(\bar z')   d\bar z' \nonumber\\
&+&{1\over 2}H \int^z\partial' X^{3}_{0,L}(z')\langle X_{0,R}^2(\bar w) X_{0,R}^2(\bar z')\rangle dz'\Biggr\rbrack_{|z|=|w|+\delta}\nonumber\\
&-&{\rm the ~other~ half ~of ~the ~commutator}\nonumber\\
&=&\Biggl\lbrack-{i\over 8}H\tilde p^3\biggl(\int^{\bar w}d\bar z' {\log(z-z')\over \bar z'}+
\int^zdz'{\log(\bar w-\bar z')\over z'}\biggr)\Biggr\rbrack_{|z|=|w|+\delta}\nonumber\\
&-&{\rm the ~other~ half ~of ~the ~commutator}
\, .
\end{eqnarray}
These integrals are evaluated in Appendix C (see, in particular, eq. (\ref{typeiv})) and so we have
\begin{eqnarray}
\Theta_{LR}^{12}(z,w,\bar z,\bar w)=
{i\pi\over 4}H\tilde p^3
\biggl(
{\varphi^2\over2 \pi}-{1\over \pi}
{\cal L}i_2(e^{i\varphi})\biggr)\, .
\end{eqnarray}

Setting $\varphi=0$, it yields
\begin{eqnarray}
\Theta_{LR}^{12}(\tau, \sigma)=- iH\tilde p^3 {\pi^2\over 12} \, .
\end{eqnarray}

Likewise, we obtain the following result for $\Theta_{RL}$,
\begin{equation}
\Theta_{RL}^{12}(z,w,\bar z,\bar w)
={i\pi\over 4}H\tilde p^3
\biggl({\varphi^2\over2 \pi}-{1\over \pi}
{\cal L}i_2(e^{i\varphi})\biggr) ,
\end{equation}
which yields for $\varphi = 0$
\begin{equation}
\Theta_{RL}^{12}(\tau, \sigma) = -iH\tilde p^3 {\pi^2\over 12} \, .
\end{equation}

Finally, adding up all four individual pieces, it comes out that the total commutator of the generalized coordinates in the
1- and 2-directions vanishes in the $H$-flux background;
in fact, it vanishes even before taking the limit $\varphi=0$. Thus, we end up with
\begin{eqnarray}
\Theta_{H}^{12}(\tau,\sigma)=\Theta_{LL}^{12}+\Theta_{RR}^{12}+\Theta_{LR}^{12}+\Theta_{RL}^{12}=0
\end{eqnarray}
and commutativity is manifest.

\subsubsection{Commutators of the geometric $f$-flux model}\label{fcommutator}

The geometric $f$-flux background is obtained from the $H$-flux face by a T-duality transformation in the 1-direction,
flipping
\begin{eqnarray}\label{tdualcal1}
\quad {\cal  X}^1_L(z,\bar z)\longrightarrow  {\cal  X}^1_L(z,\bar z)\, ,\quad {\cal  X}^1_R(z,\bar z)\longrightarrow  -{\cal  X}^1_R(z,\bar z)\,.
 \end{eqnarray}
In the computation of the corresponding commutators, we are still  using  the free fields $X^1_{0,L}(z)$ and $X^1_{0,R}(\bar z)$ of the $H$-flux
background, and, therefore, we do not have to repeat here the entire CFT computation
but we just have to implement the flip of signs of the individual commutator pieces in the appropriate way, as given by
eq. (\ref{parabolicf}). Thus, we obtain, setting $H \equiv f$,
\begin{eqnarray}
\Theta_{f}^{12}(\tau,\sigma)=\Theta_{LL}^{12}-\Theta_{RR}^{12}+\Theta_{LR}^{12}-\Theta_{RL}^{12}=0
 \end{eqnarray}
in the winding sector of string theory with non-vanishing $\tilde p^3$. Commutativity is maintained.

\subsubsection{Commutators of the non-geometric $Q$-flux model}

The non-geometric $Q$-flux background follows by performing another T-duality in the 2-direction,
\begin{eqnarray}\label{tdualcal1}
\quad {\cal  X}^2_L(z,\bar z)\longrightarrow  {\cal  X}^2_L(z,\bar z)\, ,\quad {\cal  X}^2_R(z,\bar z)\longrightarrow  -{\cal  X}^2_R(z,\bar z)\,.
 \end{eqnarray}
Now, for the first time, the total commutator is non-vanishing, since
\begin{eqnarray}
\Theta_{Q}^{12}(z, w, \bar{z}, \bar{w})&=&\Theta_{LL}^{12}+\Theta_{RR}^{12}-\Theta_{LR}^{12}-\Theta_{RL}^{12}
\nonumber\\&=&-i
\pi Q\tilde p^3
\biggl(
{\varphi^2\over2 \pi}-{1\over \pi}
 {\cal L}i_2(e^{i\varphi})\biggr)
\end{eqnarray}
with $H \equiv f \equiv Q$. Setting $\varphi=0$, it yields
\begin{eqnarray}
\Theta_{Q}^{12}(\tau,\sigma)={i\pi^2\over 3} Q\tilde p^3\, .
\end{eqnarray}

Thus, the generalized coordinates ${\cal X}^1$ and ${\cal X}^2$ of the $Q$-flux background are non-commutative.
The parameter of non-commutativity is provided by the winding number $\tilde p^3$ in the third direction of the torus.

\subsubsection{Commutators of the non-geometric $R$-flux model}

The final T-duality with respect to the 3-direction is no longer described by
the Buscher rules (\ref{tdualgb}) and (\ref{wstdual}), since the background depends explicitly on $X^3$.
An alternative prescription is provided in the language of conformal field theory
by the automorphism of the operator algebra, acting as
 \begin{equation}
 {\cal X}_L^3\,\,\rightarrow\,\,{\cal X}_L^3\, ,\quad {\cal X}_R^3\,\,\rightarrow\,\,-{\cal X}_R^3\, ,\quad \tilde{p}^3\,\,\rightarrow\,\,{p}_3\\ \, .
 \end{equation}
Thus, starting from the $H$-flux background and dualizing it in all three directions, we arrive to the non-geometric $R$-flux
background, which is completely left-right asymmetric in all three directions.
It is not anymore a geometric manifold, even in a local coordinate neighborhood.
The commutator of the generalized coordinates in the 1- and 2-directions becomes now
 \begin{eqnarray}\label{thetaR}
\Theta_{R}^{12}(\tau,\sigma)=\lbrack {\cal  X}^1(\tau,\sigma), ~ {\cal  X}^2(\tau,\sigma)\rbrack={i\pi^2\over 3}R~ p_3
\end{eqnarray}
setting $H \equiv f \equiv Q \equiv R$.

In this case, the parameter of non-commutativity is provided by the momentum number $ p_3$, which, in turn leads to
non-associativity as violation of Jacobi identity among the generalized coordinates.

This completes the derivation of the commutators in all T-dual faces of the parabolic flux model, confirming by CFT techniques the
emergence of non-commutative and/or non-associative structures in non-geometric backgrounds.


\section{Conclusions and discussion}
\setcounter{equation}{0}

In this paper we confirmed the non-commutative and non-associative algebraic structure of the parabolic flux
compactifications with constant geometric and non-geometric fluxes, using the approach of canonical T-duality in conformal field theory.
Combining this approach with the gauging procedure of Ro\v{c}ek and Verlinde, we found
that the non-commutative and non-associative relations arise in a subtle way. Namely, these structures do not refer
to the coordinates $X^I$ and the coordinates $\tilde X^I$ of the different T-dual faces of the model,
but they rather refer to generalized coordinates ${\cal X}^I$ and their dual counterparts $\tilde{\cal  X}^I$ associated to the
conserved currents of the underlying Ro\v{c}ek-Verlinde "parent" theory. The resulting commutation relations are summarized in
the table below for all T-dual faces of the parabolic flux models.

\vskip0.4cm
\begin{table}[h]
\centering
\renewcommand{\arraystretch}{1.3}
\tabcolsep10pt
\begin{tabular}{|c||c|c|}
\hline
T-dual frames  & Commutators & Three-brackets \\  \hline\hline
$H$-flux & $[\tilde{\cal X}^I, ~ \tilde{\cal X}^J] \sim H \epsilon^{IJK} \tilde{p}^K$ & $[\tilde{\cal X}^1, ~ \tilde{\cal X}^2, ~ \tilde{\cal X}^3] \sim H$\\
$f$-flux & $[{\cal X}^I, ~ \tilde{\cal X}^J] \sim f \epsilon^{IJK} \tilde{p}^K $ & $[ {\cal X}^1, ~ \tilde{\cal X}^2, ~ \tilde{\cal X}^3] \sim f$\\
$Q$-flux & $[ {\cal X}^I, ~ {\cal X}^J] \sim Q \epsilon^{IJK} \tilde{p}^K$ & $[{\cal X}^1, ~ {\cal X}^2, ~ \tilde{\cal X}^3] \sim Q$ \\
$R$-flux & $[ {\cal X}^I, ~ {\cal X}^J] \sim R \epsilon^{IJK} p^K$ & $[ {\cal X}^1, ~ {\cal X}^2, ~ {\cal X}^3] \sim R$ \\ \hline
\end{tabular}
\end{table}

\noindent
The third column of this table contains the non-trivial three-brackets among the generalized coordinates, which reflect the non-associative tri-products for each of the four flux backgrounds. This non-associative algebraic structure was further analyzed
in refs. \cite{Mylonas:2012pg, Bakas:2013jwa, Castellani:2013mka, Mylonas:2013jha, Mylonas:2014aga, Mylonas:2014kua}
in terms of 3-cocycles, star-products, tri-products and related physics topics such as membrane sigma-models,
magnetic monopole backgrounds, and their quantization.

Since the generalized coordinates ${\cal X}^I$ and their T-dual counterparts $\tilde {\cal X}^I$ do not correspond to bona-fide
conformal fields of the underlying CFT, there is no conflict with the associativity of CFT amplitudes satisfied by
on-shell physical fields. Our computations do not require staying off-shell or going on-shell, and, hence, the non-associativity of the
phase space variables $( {\cal X}^I,\tilde {\cal X}^I,p^I,\tilde p^I)$ is not forbidden on general grounds.
In our view, non-geometric flux backgrounds should fit in a broad "geometrical" picture in which the physical on-shell closed
string fields, whose operator algebra is fully associative, nevertheless live on a non-commutative and possibly non-associative phase space.
Then, as advocated in refs. \cite{Blumenhagen:2010hj,Lust:2010iy,Mylonas:2013jha}, this space possesses a minimal
volume, which is set by the $R$-flux deformation parameter together with the non-vanishing string length upon quantization.
This picture is also consistent with the role of non-associative commutation relations and tri-products in double field
theory \cite{Blumenhagen:2013zpa}.
Physical fields that satisfy the strong constraint in double field theory do not display any non-associative product structure.
Still, they can live in a non-associative doubled phase space with non-associative doubled coordinates, since there is no need for the
doubled coordinates to be physical fields satisfying the strong constraint.

It is certainly interesting to explore in more detail the relations between the doubled CFT approach of Ro\v{c}ek and Verlinde
on the string world-sheet and the doubled target space description of double field theory. In fact, it is tempting to speculate that
the "parent" world-sheet theory of Ro\v{c}ek and Verlinde is just the proper world-sheet analogue of target space double field
theory (for a different world-sheet description of double field theory see the recent work \cite{Nibbelink:2012jb,Nibbelink:2013zda}).
Then, in this context, the chirally conserved world-sheet currents ${\cal J}^I$ and $\bar {\cal J}^I$ giving rise to the generalized
coordinates ${\cal X}^I$ and $\tilde {\cal X}^I$ seem to correspond
to the doubled coordinates of double field theory. Projecting these currents to the "child" non-linear sigma-models by gauging seems to
be very similar in vain to the different ways of applying the strong constraint in double field theory.

Finally, it would be interesting to investigate other non-geometric backgrounds by similar methods, focusing, in particular,
to the novel class of non-geometric spaces that do not admit geometric duals, such as the
asymmetric orbifold conformal field theories \cite{Condeescu:2012sp,Condeescu:2013yma} and
the closely related double elliptic spaces that arose recently
by dimensional reduction of double field theory \cite{Hassler:2014sba} as well as to solutions of double field theory for
non-geometric group manifolds \cite{schulz,Blumenhagen:2014gva,Blumenhagen:2015zma}. We hope to return to these issues elsewhere.

\vskip1cm

\section*{Acknowledgements}
This work was partially supported by the ERC Advanced Grant "Strings and Gravity"
(Grant.No. 32004) and by the DFG cluster of excellence "Origin and Structure of the Universe".
This research is also implemented (I.B.) under the "ARISTEIA" action of the
operational programme education and lifelong learning and is co-funded
by the European Social Fund (ESF) and National Resources of Greece.
I.B. is grateful to the hospitality extended to him at the Arnold
Sommerfeld Center for Theoretical Physics and the Max Planck Institute for Physics in Munich during the course
of this work, whereas D.L. likes to thank the CERN theory unit for hospitality.  We thank David Andriot,
Ralph Blumenhagen, Ioannis Florakis, Falk Hassler, Olaf Hohm,
Magdalena Larfors, Peter Patalong, Felix Rennecke and Stefan Theisen for fruitful discussions.

\newpage

\appendix
\section{Monodromies of toroidal fibrations}
\setcounter{equation}{0}

The monodromies specify the gluing conditions of the fibre when going around the base space of a given fibration.
Here, in view of the applications that are discussed in this paper, we summarize the monodromy transformations of
a three-dimensional fibration with a one-dimensional circle $S^1$ with coordinate $X^3$ serving as the base and a
two-dimensional torus $T^2$  with coordinates $X^i = (X^1, ~ X^2)$ as fibre.

Let $G_{IJ}$ and $B_{IJ}$ be the components of the metric and anti-symmetric fields on $T^2$, which depend on the
base point and they are conveniently combined as $E_{IJ} = G_{IJ} + B_{IJ}$. Then, in this case,
the monodromy transformations are given in terms of $O(2,2; \mathbb{R})$ transformations, which
act on the background fields of the $T^2$ in the following way,
\begin{equation}
{ E}({X^3}+2 \pi) = {\cal M}_{O(2,2)} { E}({X^3})=\Bigl(A{E}({X^3})+B\Bigr)\Bigl(C{ E}({X^3})+D\Bigr)^{-1} ,
\label{rhseqas}
\end{equation}
by going around the base space.
Here, ${\cal M}_{O(2,2)}$ is an element of the group $O(2,2; \mathbb{R})$ represented as
\begin{equation}\label{SOdmatrix}
{\cal M}_{O(2,2)}=\begin{pmatrix} A & B \\ C & D \end{pmatrix}\, ,
\end{equation}
where $A$, $B$, $C$ and $D$ are $2$-dimensional matrices that satisfy the defining relations
\begin{equation}\label{constraints}
A^tC+C^tA=0\, ,\quad B^tD+D^tB=0\, ,\quad A^tD+C^tB= \mathbb{1}\, .
\end{equation}

The right-hand side of equation \eqn{rhseqas} provides the T-dual background $\tilde{E} (X^3)$ of $E(X^3)$ with respect
to the two commuting isometries generated by the vector fields tangent to the fibre $T^2$. Thus, when the matrix element
${\cal M} \in O(2,2; \mathbb{R})$ is not trivial and the gluing condition of the fibration involves a genuine T-duality
transformation, the space fails to be globally geometric. Likewise, if the transition function among different
patches involves a T-duality transformation that cannot be undone by a diffeomorphism, the space will fail to be
locally geometric and the notion of T-folds has to be used as substitute to ordinary differentiable manifolds.
The monodromies of different fibrations are essential for selecting the correct closed string boundary conditions
in the geometric and non-geometric spaces that are discussed in this paper.

It is convenient to parametrize the background fields of the fibre
by two complex scalars, known as the complex structure $\tau$ and the complexified
K\"ahler class $\rho$ of $T^2$, setting
\be
\tau={\frac{G_{12}}{G_{11}}}+i\ {\frac{V}{G_{11}}} ~, ~~~~~~ \rho=-B_{12}+i\ V ~,
\ee
where $V = \sqrt{{\rm det}G}$ denotes the volume element of the two-torus. Then, the $O(2,2)$ group splits naturally
as follows,
\begin{equation}
O(2,2; \mathbb{R}) \simeq SL(2, \mathbb{R})_\tau\times SL(2, \mathbb{R})_\rho \, .
\label{lourida}
\end{equation}
The group factor $SL(2, \mathbb{R})_\tau$ corresponds to reparametrization of the torus, acting as
modular transformations of its complex structure with generators $T_{\tau}: ~ \tau \rightarrow \tau + 1$ and
$S_{\tau}: ~ \tau \rightarrow -1/\tau$. It reads
\begin{equation}
\tau\rightarrow{a\tau+b\over c\tau+d}\, , \quad \textrm{with}\ \ \  ad-bc=1\,.
\end{equation}
The other factor $SL(2, \mathbb{R})_\rho$ contains the shift of the $B$-field by a constant,
$T_{\rho}: ~ \rho \rightarrow \rho+1$, as well as the T-duality transformation $S_{\rho}: ~ \rho \rightarrow -1/\rho$
that reverses the volume of the torus $V \rightarrow 1/V$ when $B_{12} = 0$. Its full action is given by
\begin{equation}
\rho\rightarrow{a'\rho+b'\over c'\rho+d'}\, ,\quad \textrm{with} \ \ \ a'd'-b'c'=1\,.
\end{equation}

The embedding of $SL(2, \mathbb{R})_\tau\times SL(2, \mathbb{R})_\rho$ in $O(2,2; \mathbb{R})$ is provided by
the following identification with the matrices $A$, $B$, $C$ and $D$ in equation (\ref{SOdmatrix}),
\begin{equation}\label{embedding}
A=a'\begin{pmatrix} a & & b \\ c & & d \end{pmatrix}\, ,\quad B=b'\begin{pmatrix} -b & & a \\ -d & & c \end{pmatrix}
\, ,\quad C=c'\begin{pmatrix} -c & -d \\ a & b \end{pmatrix}\, ,\quad D=d'\begin{pmatrix} d& -c \\ -b & a \end{pmatrix}\,.
\end{equation}
More explicitly, the matrix elements $SO(2, 2; \mathbb{R}) \simeq SL(2, \mathbb{R})_\tau\times SL(2, \mathbb{R})_\rho$
factorize as
\be
{\cal M}_{SO(2,2)} = \begin{pmatrix}  a & & b & 0 & 0 \\ c & & d & 0 & 0 \\ 0 & & 0 & d & -c \\ 0 & & 0 & -b & a \end{pmatrix}
\begin{pmatrix}  1 & 0 & 0 & 0 \\ 0 & 0 & 0 & 1 \\ 0 & 0 & 1 & 0 \\ 0 & 1 & 0 & 0 \end{pmatrix}
\begin{pmatrix}  a^{\prime} & & b^{\prime} & 0 & 0 \\ c^{\prime} & & d^{\prime} & 0 & 0 \\
0 & & 0 & d^{\prime} & -c^{\prime} \\ 0 & & 0 & -b^{\prime} & a^{\prime} \end{pmatrix}
\begin{pmatrix}  1 & 0 & 0 & 0 \\ 0 & 0 & 0 & 1 \\ 0 & 0 & 1 & 0 \\ 0 & 1 & 0 & 0 \end{pmatrix}\,,
\ee
explaining the identification \eqn{embedding}.

We also note for completeness that there are two discrete transformations that can be appended to the previous description:
the $\mathbb{Z}_2$ transformation $(\tau, ~ \rho) \rightarrow (\rho, ~ \tau)$ that exchanges complex and K\"ahler structures
and another $\mathbb{Z}_2$ transformation $(\tau, ~ \rho) \rightarrow (- \bar{\tau}, ~ - \bar{\rho})$. Both can be
intertwined with world-sheet parity.

The group elements of each $SL(2, \mathbb{R})$ factor, which are represented by the $2 \times 2$ matrices
\be
M_{\tau}=\begin{pmatrix} a & & & b \\ c & & & d \end{pmatrix} ~, ~~~~~~
M_{\rho}=\begin{pmatrix} a^{\prime} & & & b^{\prime} \\ c^{\prime} & & & d^{\prime} \end{pmatrix} ~,
\ee
are classified according to their eigenvalues into elliptic, parabolic or hyperbolic elements. Since the eigenvalues
of $M$ (it can be either $M_{\tau}$ or $M_{\rho}$) satisfy the characteristic polynomial equation
\be
\lambda^2 - {\rm tr}(M) \lambda + 1 = 0 ~,
\ee
we have the following three distinct cases:

\noindent
\begin{itemize}
\item If $| {\rm tr}(M) | < 2$, the two eigenvalues will be complex conjugate and $M$ is called elliptic element.
\item If $| {\rm tr}(M) | = 2$, the two eigenvalues will be real and equal and $M$ is called parabolic element.
\item If $| {\rm tr}(M) | > 2$, the two eigenvalues will be real and unequal and $M$ is called hyperbolic element.
\end{itemize}

Elliptic group elements are characterized by the property $M^n = \mathbb{1}$ for some finite integer $n$,
like
\be
M = \begin{pmatrix}  0 &  & 1 \\ -1 & & 0 \end{pmatrix} ,
\ee
which has $M^4 = \mathbb{1}$, whereas parabolic elements, like
\be
M = \begin{pmatrix}  1 &  & 1 \\ 0 & & 1 \end{pmatrix} ,
\ee
do not share this property.

In this paper we are only concerned with parabolic group elements $M$, and, for that reason, the corresponding
closed string backgrounds associated to $T^2$ fibrations over $S^1$ with constant flux are called parabolic.

\section{Parabolic backgrounds and monodromies of currents}
\setcounter{equation}{0}

In this appendix we give a brief account of the toroidal vacua of closed string theory with constant
fluxes and determine the monodromy properties of their currents in geometric as well as in non-geometric faces.

\subsection{The parabolic backgrounds with constant flux}

Recall that the constant flux backgrounds in three dimensions come in four different versions, which are related to each other by
successive T-duality transformations.
Concretely, we consider the compactification on a three-dimensional
torus $T^3$, which can be viewed as an ${\cal F}=T^{n}$ fibration  over the $(3-n)$-dimensional base space
${\cal B}=T^{3-n}$ for $n=0, \, 1, \, 2, \, 3$. In the following, we consider all four different cases.
The starting point, provided by $n=0$,
is a flat three-torus $T^3$ parametrized by $X^{I=1,2,3}$ with periodic identifications $X^I \sim X^I+ 2 \pi r^I$,
where $r^I$ are the three radii of $T^3$, which are all set equal to $1$ for convenience. There is also a 3-form
$H$-flux $H_3=H \, d X^1 \wedge d X^2 \wedge d X^3$ with constant $H$.
The metric and the corresponding $B$-field are taken to be
\begin{equation} \label{Abfield}
G_{IJ}=\delta_{IJ}\, ,\quad B_{12} = {H} X^3 \ , \quad B_{13} = 0 = B_{23} \, .
\end{equation}

The $H$-flux has to obey a topological quantization condition, which is given, in general, by (see, for instance, \cite{Grana:2005jc})
\begin{equation}
{1\over 4\pi^2 \alpha^{\prime}}\int_{T^3} H=k\, ,\quad k\in{\mathbb Z}\, .
\end{equation}
Since the volume of the unit three-torus is $(2\pi)^3$, it follows immediately that the quantization condition for $H$ takes the
following form, in units $\alpha^{\prime} =1$,
\begin{equation}
 H={k\over 2 \pi}\, ,\quad k\in{\mathbb Z}\, .
\label{koukliaw}
\end{equation}
For the discussion of T-duality transformations as well as for the CFT computation of the commutators, we consider the simplest case $k=1$.

The quantization condition \eqn{koukliaw} appears to be incompatible with the dilute flux approximation used in the definition of 
$CFT_{\rm H}$, but this is illusive because we have normalized the radii of $T^3$ as well as $\alpha^{\prime}$ to 1 for notational convenience. 
$H$ has units $[L]^{-1}$ and it can be made very small for appropriate choice of radii as can be seen by reinstating the parameters 
in natural units in which $\alpha^{\prime} = l_s^2$.

Performing a T-duality transformation on the one-dimensional circle fibre ${\cal F}=T^{1}_{X^1}$ with coordinate $X^1$,
one obtains the Heisenberg nilmanifold,
which is a twisted torus without $B$-field corresponding to the case $n=1$. This background is topologically distinct from $T^3$,
since it has a different homology group.
Its Levi-Civita connection is related to the structure constants $f^I_{JK}$ of the Heisenberg group, which are the geometric fluxes.
We also note in this respect that the flat 3-torus corresponds to Bianchi-I geometry, whereas the twisted torus to Bianchi-II.
The background fields of the twisted torus have the following form, setting $H\equiv f$,
\begin{equation}
G  = \begin{pmatrix} 1 & {fX^3} & 0 \\  {fX^3} & 1 + \left({fX^3}\right)^2 & 0 \\ 0 & 0 & 1 \end{pmatrix}\, ,\quad B=0 \ .\label{metric}
\end{equation}

Performing another T-duality in the $X^2$-direction of the two-dimensional torus fibre
${\cal F}=T^{2}_{X_1,X_2}$, one obtains the so-called $Q$-space that corresponds to the case $n=2$.
This background is again a $T^2$-fibration, but the corresponding metric and $B$-field are defined only locally
and not globally. Setting $H\equiv f\equiv Q$ for notational purposes, $G$ and $B$ take the form
\begin{equation}
G = g (X^3)\begin{pmatrix}  1 & 0 & 0 \\ 0 & 1 & 0 \\ 0 & 0 & \frac{1}{g(X^3)} \end{pmatrix}\ ,
\ B = g(X^3) \begin{pmatrix} 0 &- QX^3 & 0 \\ QX^3 & 0 & 0 \\ 0 & 0 & 0 \end{pmatrix}\ ,
\ g(X^3)={1 \over 1-\left({Q X^3} \right)^2} \, .\label{eq:nongeofields}
\end{equation}
The $Q$-space is not a Riemannian manifold because the fibre ${\cal F}$ is glued with a T-duality transformation
when transporting it once around the base ${\cal B}$ and not with a standard diffeomorphism.

The three backgrounds we have discussed so far take particularly simple form to linear order in the fluxes,
summarized in the table below
\begin{center}
\begin{tabular}{|c||c|}
\hline
Backgrounds &  Target space fields \\
\hline
\hline
Torus + $H$-flux &  $G = \begin{pmatrix} 1 & 0 & 0 \\ 0 & 1 & 0 \\ 0 & 0 & 1 \end{pmatrix}\ ,
\ B = \begin{pmatrix} 0 & {H X^3} & 0 \\ - {H X^3} & 0 & 0 \\ 0 & 0 & 0 \end{pmatrix} $ \\
\hline
Twisted torus &  $G = \begin{pmatrix} 1 &  {fX^3} & 0 \\  {fX^3} & 1 & 0 \\ 0 & 0 & 1 \end{pmatrix} + {\cal O}(f^2)\ , \ B = 0 $ \\
\hline
Non-geom. &  $G = \begin{pmatrix}  1 & 0 & 0 \\ 0 & 1 & 0 \\ 0 & 0 & 1 \end{pmatrix} + {\cal O}(Q^2)\ , \ B =  \begin{pmatrix} 0 & -{Q}X^3 & 0 \\ {Q}X^3 & 0 & 0 \\ 0 & 0 & 0 \end{pmatrix} + {\cal O}(Q^2)$  \\
\hline
\end{tabular}
\end{center}
Thus, to linear order, one sees that the $Q$-flux background looks locally identical to the $H$-flux background, up to a sign
flip in the flux. However, these two background are not the same globally, since the
$Q$-flux background is glued together utilizing a $T$-duality transformation.

Finally, one can also consider applying a T-duality transformation to the entire three-dimensional torus, which is
seen as a fibration with fibre ${\cal F}=T^{3}_{X_1,X_2,X_3}$ over a base point corresponding to the case $n=3$.
This last T-duality looks somewhat problematic, since the $X_3$-direction is no longer a Killing isometry of the background
and the standard Buscher rules cannot be applied as they stand. Nevertheless,
this final step can be performed in the context of the underlying CFT by making a sign flip in the corresponding coordinate,
hereby mimicking the action of T-duality on the currents, ending up with a "space" that is left-right
asymmetric in all three directions. This defines the so called $R$-space, setting $H\equiv f\equiv Q \equiv R$
for notational purposes, which is non-geometric locally as well as globally.

\subsection{Monodromies of geometric $H$-flux model}

The monodromies of the original $H$-flux background are given by shifts of the $B$-field, which act on $\tau$ and $\rho$
in the following way under $X^3\rightarrow X^3+2\pi$,
\begin{equation}
\rho(X^3+2\pi)=\rho(X^3)-{\pi} H\, , ~~~~~~  \tau(X^3+2\pi)=\tau(X^3)\, .
\end{equation}
This gauge transformation of the $B$-field corresponds to the $O(2,2)$ monodromy transformation in the $1, \, 2$-directions
given by the group elements
\be\label{bfieldodd}
{\cal M}_{O(2,2)} \ = \
\begin{pmatrix}1&0&0&{\pi} H\\ 0&1&-{\pi} H&0\\0&0&1&0\\0&0&0&1 \end{pmatrix}\  ,
\ee
which indeed act on the K\"ahler and complex moduli $\rho$ and $\tau$ as above.

The shift of the base coordinate $X^3\rightarrow X^3+2\pi$ induces the following closed string boundary condition
on the left- and right-moving closed string coordinates $X^3_L(z)$ and $X^3_R(\bar z)$,
\begin{equation}
X_L^3(e^{2i\pi }z)=X_L^3(z)-\pi \tilde p^3\, ,\quad X_R^3(e^{-2i\pi }\bar z)=X_R^3(\bar z)-\pi \tilde p^3\, ,
\end{equation}
where $\tilde p^3=p_L^3-p_R^3$ is the winding number of the closed string in the third direction.
Then, using the winding shifts of $X^3_{L,R}$, it follows from eq.(\ref{q1all}) that
the associated closed string boundary conditions of the CFT currents ${\cal J}^I$ and $\bar{\cal J}^I$ to linear order in the flux are
\begin{eqnarray}
{\cal  J}^1 (e^{-2i\pi }z,e^{2i\pi}\bar z)&= & {\cal  J}^1 (z,\bar z)-{\pi\over 2} H \tilde p^3 { \cal J}^2 (z,\bar z)    \, ,\nonumber\\
\bar { \cal J}^1 (e^{-2i\pi }z,e^{2i\pi}\bar z) &= &\bar { \cal J}^1 (z,\bar z)+{\pi\over 2} H \tilde p^3 \bar{\cal  J}^2 (z,\bar z)  \, , \nonumber\\
 { \cal J}^2 (e^{-2i\pi }z,e^{2i\pi}\bar z) &=&  { \cal J}^2 (z,\bar z)+{\pi\over 2} H \tilde p^3 {\cal  J}^1 (z,\bar z)  \, ,\nonumber\\
\bar {\cal  J}^2 (e^{-2i\pi }z,e^{2i\pi}\bar z) &= &\bar { \cal J}^2 (z,\bar z) -{\pi\over 2}H \tilde p^3\bar {\cal  J}^1 (z,\bar z)     \; ,\nonumber\\
 { \cal J}^3 (e^{-2i\pi }z,e^{2i\pi}\bar z)& =& {\cal  J}^3 (z,\bar z)  \, ,\nonumber\\
\bar {\cal  J}^3 (e^{-2i\pi }z,e^{2i\pi}\bar z)& =&\bar { \cal J}^3 (z,\bar z)
\; .\label{currentmonod}
\end{eqnarray}
Thus, the monodromies of the $H$-flux background act in a left-right asymmetric way on the currents,
transforming  the currents into the dual currents and vice versa.

\subsection{Monodromies of geometric $f$-flux model}

Perform a T-duality transformation in the $1$-direction we obtain the parabolic $f$-flux background with geometric flux.
The corresponding monodromies are given as shifts of the complex structure $\tau$ under $X^3\rightarrow X^3+2\pi$,
\begin{equation}
\rho(X^3+2\pi)=\rho(X^3)\, , ~~~~~~  \tau(X^3+2\pi)=\tau(X^3)-{\pi} f\, ,
\end{equation}
associated to the $O(2,2)$ monodromy transformation in the $1, \, 2$-directions with group elements
\be\label{bfieldodd}
{\cal M}_{O(2,2)} \ = \
\begin{pmatrix}1&-{\pi} f&0&0\\ 0&1&0&0\\0&0&1&0\\0&0&{\pi} f&1 \end{pmatrix}\, .
\ee
Since T-duality acts in this case as $\bar { \cal J}^1\rightarrow -\bar { \cal J}^1$, the corresponding closed
string boundary conditions of the currents ${\cal J}^I$ take the following form,
\begin{eqnarray}
{ \cal J}^1 (e^{-2i\pi }z,e^{2i\pi}\bar z)&= & { \cal J}^1 (z,\bar z)-{\pi\over 2} f \tilde p^3 {\cal  J}^2 (z,\bar z)    \, ,\nonumber\\
\bar { \cal J}^1 (e^{-2i\pi }z,e^{2i\pi}\bar z) &= &\bar { \cal J}^1 (z,\bar z)-{\pi\over 2} f \tilde p^3 \bar { \cal J}^2 (z,\bar z)  \, , \nonumber\\
 {\cal  J}^2 (e^{-2i\pi }z,e^{2i\pi}\bar z) &=&  { \cal J}^2 (z,\bar z)+{\pi\over 2} f \tilde p^3 { \cal J}^1 (z,\bar z)  \, ,\nonumber\\
\bar { \cal J}^2 (e^{-2i\pi }z,e^{2i\pi}\bar z) &= &\bar { \cal J}^2 (z,\bar z) +{\pi\over 2} f \tilde p^3 \bar{ \cal J}^1 (z,\bar z)     \; ,\nonumber\\
 { \cal J}^3 (e^{-2i\pi }z,e^{2i\pi}\bar z)& =& { \cal J}^3 (z,\bar z)  \, ,\nonumber\\
\bar { \cal J}^3 (e^{-2i\pi }z,e^{2i\pi}\bar z)& =&\bar { \cal J}^3 (z,\bar z)
\; .\label{currentfmonod}
\end{eqnarray}
Thus, for the $f$-flux background, the monodromies act in left-right symmetric way on the currents, meaning that the
currents and the dual currents do not mix with each other.

\subsection{Monodromies of non-geometric $Q$-flux model}

Next, performing another T-duality in the $2$-direction brings us to the non-geometric background with $Q$-flux.
The monodromy of the torus is now determined by the following transformation of $\rho$ and $\tau$ under $X^3\rightarrow X^3+2\pi$,
\be\label{nonlinrho}
\rho(X^3+2\pi) \ = \ \frac{\rho}{1-{\pi}Q \rho(X^3)} \, , ~~~~~~ \tau(X^3+2\pi)=\tau(X^3)\, ,
\ee
which is an $O(2,2)$ transformation in the $1, \, 2$-directions with group elements
\be\label{O22Hthird}
{\cal M}_{O(2,2)} \ = \
\begin{pmatrix}1&0&0&0\\ 0&1&0&0\\0&{\pi}Q&1&0\\-{\pi}Q&0&0&1 \end{pmatrix}\  .
\ee
Here, T-duality is performed by flipping the signs $\bar { \cal J}^1\rightarrow -\bar { \cal J}^1$ and
$\bar { \cal J}^2\rightarrow -\bar { \cal J}^2$,
and, thus, the closed string boundary conditions of the currents are taking the form
\begin{eqnarray}
{\cal  J}^1 (e^{-2i\pi }z,e^{2i\pi}\bar z)&= & { \cal J}^1 (z,\bar z)-{\pi\over 2} Q \tilde p^3 {\cal  J}^2 (z,\bar z)    \, ,\nonumber\\
\bar {\cal  J}^1 (e^{-2i\pi }z,e^{2i\pi}\bar z) &= &\bar { \cal J}^1 (z,\bar z)+{\pi\over 2} Q \tilde p^3 \bar { \cal J}^2 (z,\bar z)  \, , \nonumber\\
 { \cal J}^2 (e^{-2i\pi }z,e^{2i\pi}\bar z) &=&  { \cal J}^2 (z,\bar z)+{\pi\over2} Q \tilde p^3 { \cal J}^1 (z,\bar z)  \, ,\nonumber\\
\bar {\cal  J}^2 (e^{-2i\pi }z,e^{2i\pi}\bar z) &= &\bar {\cal  J}^2 (z,\bar z) -{\pi\over2 } Q \tilde p^3 \bar { \cal J}^1 (z,\bar z)     \; ,\nonumber\\
 { \cal J}^3 (e^{-2i\pi }z,e^{2i\pi}\bar z)& =& { \cal J}^3 (z,\bar z)  \, ,\nonumber\\
\bar {\cal  J}^3 (e^{-2i\pi }z,e^{2i\pi}\bar z)& =&\bar {\cal  J}^3 (z,\bar z)
\; .\label{currentQmonod}
\end{eqnarray}
The monodromies are again left-right asymmetric, as for the $H$-flux model. In fact, they agree with the monodromies of
$H$-flux background to linear order in the fluxes, up to a minus sign in the flux, as expected from the previous general
discussion.

\subsection{Monodromies of non-geometric $R$-flux model}

Finally, coming to the $R$-flux model, the closed string boundary conditions for the currents ${\cal J}^I$ are
determined by the momentum number $p_3$ and they take the following form,
 \begin{eqnarray}
{ \cal J}^1 (e^{-2i\pi }z,e^{2i\pi}\bar z)&= & { \cal J}^1 (z,\bar z)-{\pi\over2} R p_3 {\cal  J}^2 (z,\bar z)    \, ,\nonumber\\
\bar { \cal J}^1 (e^{-2i\pi }z,e^{2i\pi}\bar z) &= &\bar { \cal J}^1 (z,\bar z)+{\pi\over2} R p_3 \bar {\cal  J}^2 (z,\bar z)  \, , \nonumber\\
 { \cal J}^2 (e^{-2i\pi }z,e^{2i\pi}\bar z) &=&  {\cal  J}^2 (z,\bar z)+{\pi\over2} R p_3 { \cal J}^1 (z,\bar z)  \, ,\nonumber\\
\bar { \cal J}^2 (e^{-2i\pi }z,e^{2i\pi}\bar z) &= &\bar {\cal  J}^2 (z,\bar z) -{\pi\over2} R p_3 \bar { \cal J}^1 (z,\bar z)     \; ,\nonumber\\
 {\cal  J}^3 (e^{-2i\pi }z,e^{2i\pi}\bar z)& =& {\cal  J}^3 (z,\bar z)  \, ,\nonumber\\
\bar {\cal  J}^3 (e^{-2i\pi }z,e^{2i\pi}\bar z)& =&\bar { \cal J}^3 (z,\bar z)
\; . \label{currentRmonod}
\end{eqnarray}

This concludes the description of how the monodromies act on the currents in all T-dual faces of the toroidal flux model.
Judging from the action on the $\rho$ and $\tau$ moduli, the corresponding $SL(2, \mathbb{R})$ group elements are parabolic
(i.e., of infinite order) using the nomenclature of Appendix A.

Summarizing, the chain of T-duality transformations of the toroidal background with constant $H$-flux, encoded in the diagram
\begin{equation}\label{eq:TdualityChain}
H_{IJK} \stackrel{T_{I}}{\longrightarrow} f^{I}_{JK}
\stackrel{T_{J}}{\longrightarrow} Q_{K}^{IJ}
\stackrel{T_{K}}{\longrightarrow} R^{IJK}\, ; ~~~~~~ I, \, J, \, K=1,\, \dots , \, 3\, ,
\end{equation}
provide the parabolic flux models of closed strings that are in focus in the present work.

\section{Some relevant dilogarithmic integrals}
\setcounter{equation}{0}

In this appendix we collect some useful expressions involving the dilogarithm function based on series
expansions and integral formulae (see, for instance, \cite{zagier}). We also evaluate certain commutators
used in the main text based on the dilogarithm function.

\subsection{Delta-function, step-function and the logarithm}

The delta-function on a circle of unit radius is defined through the infinite series
\begin{equation}
\delta(\varphi)=\sum_{m\in{\cal Z}} e^{im\varphi}\, .
 \end{equation}
Likewise, the step-function, which is the integral of the delta-function,
\be
\delta(\varphi)={1\over 2}{d\over d\varphi}\epsilon(\varphi)\, ,
\ee
is represented by the infinite series
\begin{equation}\label{stepdelta}
\epsilon(\varphi)= -{i\over \pi}\sum_{m\neq 0} {1\over m}e^{im\varphi}+{\varphi\over \pi}
\end{equation}
so that  $\epsilon(\varphi)=1$ for $\varphi>0$ (i.e., for $\sigma>\sigma'$), $\epsilon(\varphi)=-1$ for $\varphi<0$
(i.e., for $\sigma<\sigma'$) and  $\epsilon(\varphi)=0$ for $\varphi=0$ (i.e., for $\sigma=\sigma'$).
Finally, the logarithm is represented by the infinite series
\begin{equation}
\log(1-e^{i\varphi})=-\sum_{m=1}^\infty {1\over m}e^{im\varphi}\, .
\end{equation}

Hence, one obtains the following relation between the logarithm, the step-function and the delta-function,
\begin{eqnarray}
\epsilon(\varphi)&=&{i\over \pi}\log|1-e^{i\varphi}|^2+{\varphi\over \pi}\, ,\nonumber\\
\delta(\varphi)&=&{i\over 2\pi}{d\over d\varphi}\log|1-e^{i\varphi}|^2+{1\over2 \pi}\nonumber\\
&=&-{1\over 2\pi}\biggl({e^{i\varphi}\over1-e^{i\varphi}}+{e^{-i\varphi}\over1-e^{-i\varphi}}-1\biggr)\, .
\end{eqnarray}

It useful to introduce a regulator $e^{-\delta}$ with $\delta >0$ when writing infinite series,
as in the example
\begin{equation}
 \lim_{\delta\rightarrow 0_+}{1\over 1-e^{-\delta+i\varphi}}=\sum_{m=0}^\infty e^{im\varphi}\, .
\end{equation}
The $\delta$-prescription is implicitly assumed hereafter.

 \subsection{The dilogarithm function}

The dilogarithm function follows by integrating the logarithm through the defining relation
 \be
 \log(1-e^{i\varphi}) = i{d\over d\varphi}Li_2(e^{i\varphi})\, ,
 \ee
leading to the expression
\begin{eqnarray}
 Li_2(e^{i\varphi})&=&-i\int^\varphi d\tilde\varphi\log(1-e^{i\tilde\varphi}) \nonumber\\
 &=&\sum_{m=1}^\infty {1\over m^2}e^{im\varphi}\, .
 \end{eqnarray}
When these relations are written in terms of the complex variable $z=e^{i\varphi}$ they assume the form
 \begin{eqnarray}
 Li_2(z)&=&-i\int^z_0 d\tilde z{\log(1-\tilde z)\over\tilde z} \nonumber\\
 &=&\sum_{m=1}^\infty {1\over m^2}z^{m}\, .
 \end{eqnarray}

Finally, we define the closely related function ${\cal L}i_2(e^{i\varphi})$ as follows
 \begin{eqnarray}
 {\cal L}i_2(e^{i\varphi})&=&Li_2(e^{i\varphi})+Li_2(e^{-i\varphi})=
-i\int^\varphi d\tilde\varphi\log\biggl({1-e^{i\tilde\varphi}\over  1-e^{-i\tilde\varphi}   }\biggr) \nonumber\\
&=& \sum_{m\neq 0}^\infty {1\over m^2}e^{im\varphi}\, .
 \end{eqnarray}

\subsection{Certain dilogarithmic commutators}

We are now in position to evaluate certain commutators that appear in the main text and help to establish the
non-commutative structure of the non-geometric flux backgrounds.
First consider the following (prototype) commutator

\vskip0.3cm
\noindent
{\bf (i) $\lbrack   X(z), ~ \int^wd\tilde w{X(\tilde w)\over \tilde w}\rbrack$: }
\vskip0.2cm

\noindent
This example is used as prototype for the computation of all other commutators in the following. Using the
$\delta$-prescription, it assumes the form
\begin{eqnarray}
&{~}&\biggl\lbrack X(z), ~ \int^wd\tilde w{X(\tilde w)\over \tilde w}\biggr\rbrack_{|z|=|w|}=
\nonumber \\ & &
{1\over 2} \lim_{\delta\rightarrow 0_+} \biggl(  \int^wd\tilde w{\log({z-\tilde w})\over \tilde w}_{|z|=|w|+\delta}-\int^wd\tilde w{\log({z-\tilde w})\over \tilde w}_{|z|=|w|-\delta} \biggr) .
\end{eqnarray}
Choosing $z=e^{\pm\delta-i\sigma}$, $w=e^{-i\sigma'}$ and $\tilde w=e^{-i\tilde \sigma'}$ and setting $\varphi=\sigma-\sigma'$ and  $\tilde\varphi=\sigma-\tilde\sigma'$, we note that integration takes place over a circular path of unit modulus.
Since $d\tilde w=i\tilde wd\tilde \varphi$, the expression above becomes
\begin{eqnarray}
\biggl\lbrack X(z), ~ \int^wd\tilde w{X(\tilde w)\over \tilde w}\biggr\rbrack_{|z|=|w|}&=&
{\pi\over 2}\int^{\varphi} d\tilde\varphi\biggl(
1+{\tilde\varphi\over \pi}-{i\over \pi}
\sum_{m\neq 0}{1\over m} e^{im\tilde\varphi}\biggr)
\nonumber\\
&=&{\pi\over 2}\biggl(
\varphi+{\varphi^2\over2 \pi}-{1\over \pi}
\sum_{m\neq 0}{1\over m^2} e^{im \varphi}\biggr)
\nonumber\\
&=&{\pi\over 2}\biggl(
\varphi+{\varphi^2\over2 \pi}-{1\over \pi}
Li_2(e^{i\varphi})-{1\over \pi}
Li_2(e^{-i\varphi})\biggr)\nonumber\\
&=&{\pi\over 2}\biggl(
\varphi+{\varphi^2\over2 \pi}-{1\over \pi}
{\cal L}i_2(e^{i\varphi})
\biggr)
\, .
\end{eqnarray}

Thus, for $\varphi=0$, we obtain the result
\begin{eqnarray}
\biggl\lbrack X(z), ~ \int^wd\tilde w{X(\tilde w)\over \tilde w}\biggr\rbrack_{|z|=|w|}=-{\pi^2\over 6}\, .
\end{eqnarray}

\vskip0.3cm
\noindent
{\bf (ii) $\lbrack   X(z), ~ \int^wd\tilde w\partial_{\tilde w}X(\tilde w)\log \tilde w\rbrack$: }
\vskip0.2cm

\noindent
Using the $\delta$-prescription, as before, this particular commutator is given by
\begin{eqnarray}
&{~}&\biggl\lbrack
X(z), ~ \int^wd\tilde w\partial_{\tilde w}X(\tilde w)\log \tilde w
\biggr\rbrack_{|z|=|w|}=
\nonumber \\ &=&-
{1\over 2} \lim_{\delta\rightarrow 0_+} \biggl(  \int^wd\tilde w{1\over z-\tilde w}\log\tilde w_{|z|=|w|+\delta}-
\int^wd\tilde w{1\over z-\tilde w}\log\tilde w_{|z|=|w|-\delta} \biggr) .
\end{eqnarray}
Integration by parts bring it into the form of the previous example, noting that the boundary contributions arising from partial
integration cancel from the two terms of the commutator. Thus, we obtain the result
\begin{eqnarray}
&{~}&\biggl\lbrack
X(z), ~ \int^wd\tilde w\partial_{\tilde w}X(\tilde w)\log \tilde w
\biggr\rbrack_{|z|=|w|}=
\nonumber \\ &=&
-{1\over 2} \lim_{\delta\rightarrow 0_+} \biggl(  \int^wd\tilde w{\log({z-\tilde w})\over \tilde w}_{|z|=|w|+\delta}-\int^wd\tilde w{\log({z-\tilde w})\over \tilde w}_{|z|=|w|-\delta}
\biggr)\nonumber\\
&=&
-{\pi\over 2}\biggl(
\varphi+{\varphi^2\over2 \pi}-{1\over \pi}
 {\cal L}i_2(e^{i\varphi})\biggr)\, .
\end{eqnarray}

\vskip0.3cm
\noindent
{\bf (iii) $\lbrack   X(z), ~ \int^wd\tilde w\partial_{\tilde w}X(\tilde w)\log \bar{\tilde w}\rbrack$: }
\vskip0.2cm

\noindent
This example as well as the next one involve non-homomorphic expressions.
This particular commutator is relevant for computing $\Theta_{LL}$ in the main text, whereas its complex conjugate expression
is relevant one for computing $\Theta_{RR}$.
Since we are always integrating over circular paths, we use $\log \bar{\tilde w}=-\log {\tilde w}$ to cast this commutator
into the form
\begin{eqnarray}\label{typeiii}
&{~}&\biggl\lbrack
X(z), ~ \int^wd\tilde w\partial_{\tilde w}X(\tilde w)\log\bar {\tilde w}
\biggr\rbrack_{|z|=|w|}=
\nonumber \\ &=&
{1\over 2} \lim_{\delta\rightarrow 0_+} \biggl(  \int^wd\tilde w{1\over z-\tilde w}\log\tilde w_{|z|=|w|+\delta}-\int^wd\tilde w{1\over z-\tilde w}\log\tilde w_{|z|=|w|-\delta}
\biggr)\nonumber\\
&=&
{\pi\over 2}\biggl(
\varphi+{\varphi^2\over2 \pi}-{1\over \pi}
{\cal L}i_2(e^{i\varphi})\biggr)\, .
\end{eqnarray}

\vskip0.3cm
\noindent
{\bf (iv) $\lbrack   X(z), ~ \int^{\bar w}d\bar {\tilde w}{X(\tilde w)\over \bar{ \tilde w}}\rbrack$: }
\vskip0.2cm

\noindent
This last example is relevant for computing $\Theta_{LR}$ and $\Theta_{RL}$ in the main text.
As before, we have
\begin{eqnarray}
&{~}&\biggl\lbrack X(z), ~ \int^{\bar w}d\bar{\tilde w}{X( {\tilde w})\over\bar{ \tilde w}}\biggr\rbrack_{|z|=|w|} =
\nonumber \\ & &
{1\over 2} \lim_{\delta\rightarrow 0_+} \biggl(  \int^{\bar w}d\bar {\tilde w}{\log({z-{\tilde w}})\over \bar{\tilde w}}_{|z|=|w|+\delta}-\int^{\bar w}d\bar{\tilde w}
{\log({z-\tilde w})\over \bar{\tilde w}}_{|z|=|w|-\delta}
\biggr)
\end{eqnarray}

Since $d\bar {\tilde w}=-i\bar{ \tilde w}d\tilde \varphi$, it follows that
\begin{eqnarray}\label{typeiv}
\biggl\lbrack X(z),~ \int^{\bar w}d\bar{\tilde w}{X( {\tilde w})\over\bar{ \tilde w}}\biggr\rbrack_{|z|=|w|}
&=&-{\pi\over 2}\int d\tilde\varphi\biggl(
1+{\tilde\varphi\over \pi}-{i\over \pi}
\sum_{m\neq 0}{1\over m} e^{im\tilde\varphi}\biggr)
\nonumber\\
&=&-{\pi\over 2}\biggl(
\varphi+{\varphi^2\over2 \pi}-{1\over \pi}
\sum_{m\neq 0}{1\over m^2} e^{im \varphi}\biggr)
\nonumber\\
&=&-{\pi\over 2}\biggl(
\varphi+{\varphi^2\over2 \pi}-{1\over \pi}
{\cal L}i_2(e^{i\varphi})\biggr)\, .
\end{eqnarray}

\end{document}